\documentclass[sigconf, nonacm]{acmart}

\usepackage[ruled, vlined]{algorithm2e}
\usepackage{subfigure}
\usepackage{multirow}
\usepackage{balance}
\usepackage[normalem]{ulem}
\usepackage{amsmath,bm}
\usepackage{threeparttable}
\usepackage{color}
\usepackage{url}
\usepackage{appendix}

\newcommand\vldbdoi{10.14778/3489496.3489504}
\newcommand\vldbpages{237-245}
\newcommand\vldbvolume{15}
\newcommand\vldbissue{2}
\newcommand\vldbyear{2022}

\newcommand\vldbtitle{\shorttitle}
\newcommand\vldbavailabilityurl{}
\newcommand\vldbpagestyle{empty}

\begin{document}
\title{LargeEA: Aligning Entities for Large-scale Knowledge Graphs}


\author{Congcong Ge, Xiaoze Liu, Lu Chen, Yunjun Gao}
\affiliation{%
  \institution{College of Computer Science}
  \city{Zhejiang University, Hangzhou}
  \state{China}
}
\email{{gcc, xiaoze, luchen, gaoyj}@zju.edu.cn}

\author{Baihua Zheng}
\affiliation{%
  \institution{School of Computing and Information Systems}
  \city{Singapore Management University}
  \state{Singapore}
}
\email{bhzheng@smu.edu.sg}

\begin{abstract}
Entity alignment (EA) aims to find equivalent entities in different knowledge graphs (KGs).
Current EA approaches suffer from scalability issues, limiting their usage in real-world EA scenarios.
To tackle this challenge, we propose \textsf{LargeEA} to align entities between large-scale KGs.
\textsf{LargeEA} consists of two channels, i.e., structure channel and name channel.
For the structure channel, we present METIS-CPS, a memory-saving mini-batch generation strategy, to partition large KGs into smaller mini-batches.
\textsf{LargeEA}, designed as a general tool, can adopt any existing EA approach to learn entities' structural features within each mini-batch independently.
For the name channel, we first introduce NFF, a name feature fusion method, to capture rich name features of entities without involving any complex training process;
we then exploit a name-based data augmentation to generate seed alignment without any human intervention.
Such design fits common real-world scenarios much better, as seed alignment is not always available.
Finally, \textsf{LargeEA} derives the EA results by fusing the structural features and name features of entities.
Since no widely-acknowledged benchmark is available for large-scale EA evaluation, we also develop a large-scale EA benchmark called DBP1M extracted from real-world KGs.
Extensive experiments confirm the superiority of \textsf{LargeEA} against state-of-the-art competitors.
\end{abstract}

\maketitle

\pagestyle{\vldbpagestyle}
\begingroup\small\noindent\raggedright\textbf{PVLDB Reference Format:}\\
{Congcong Ge, Xiaoze Liu, Lu Chen, Baihua Zheng, Yunjun Gao.} \vldbtitle. PVLDB, \vldbvolume(\vldbissue): \vldbpages, \vldbyear.\\
\href{https://doi.org/\vldbdoi}{doi:\vldbdoi}
\endgroup
\begingroup
\renewcommand\thefootnote{}\footnote{\noindent
This work is licensed under the Creative Commons BY-NC-ND 4.0 International License. Visit \url{https://creativecommons.org/licenses/by-nc-nd/4.0/} to view a copy of this license. For any use beyond those covered by this license, obtain permission by emailing \href{mailto:info@vldb.org}{info@vldb.org}. Copyright is held by the owner/author(s). Publication rights licensed to the VLDB Endowment. \\
\raggedright Proceedings of the VLDB Endowment, Vol. \vldbvolume, No. \vldbissue\ %
ISSN 2150-8097. \\
\href{https://doi.org/\vldbdoi}{doi:\vldbdoi} \\
}\addtocounter{footnote}{-1}\endgroup

\ifdefempty{\vldbavailabilityurl}{}{
\vspace{.3cm}
\begingroup\small\noindent\raggedright\textbf{PVLDB Artifact Availability:}\\
The source code, data, and/or other artifacts have been made available at \url{\vldbavailabilityurl}.
\endgroup
}

\section{Introduction}
\label{sec:intro}

A knowledge graph (KG) consists of various entities and relations.
KGs are the backbone of many real-world knowledge-driven applications, such as semantic search~\cite{XiongPC17} and recommendation systems~\cite{ZhangYLXM2016}.
Since real-world KGs (e.g., YAGO3~\cite{YAGO3}) are known to be highly incomplete,
how to expand KGs to improve the quality of the knowledge-driven applications becomes increasingly important.
EA is a prerequisite for expanding the coverage
of a unified KG.
It aims to find entities from two KGs that refer to the same real-world object, according to the following three steps: (i) taking two input KGs and collecting \emph{seed alignment}; (ii) training an EA model guided by the seed alignment; and (iii) aligning the equivalent entities between the two input KGs based on the trained model.

Existing solutions to EA mainly rely on the structural features of entities~\cite{RREA20, AliNet20, HyperKA20}.
They assume that the neighbors of two equivalent entities in KGs are equivalent as well~\cite{AttrGNN20}.
Besides, recent studies have shown that incorporating \emph{side information} of KGs (e.g., entity names~\cite{DegreeAware20}) facilitates the structure-based EA~\cite{PARIS2011VLDB, AttrGNN20},
as equivalent entities usually have similar side information.
Existing EA methods have demonstrated considerable performance on several representative benchmarks (such as DBP15K~\cite{JAPE17} and IDS100K~\cite{OpenEA2020VLDB}).
However, we find out that they suffer from scalability issues.
They cannot effectively align entities in real-life KGs that are much larger than the existing benchmarks. For example,
%
%
among all the popular EA benchmarks, the largest KG contains only $100,000$ entities.
\cite{tkdeSurvey20} has indicated that current EA methods, when handling $100,000$ entities, either (i) require a huge memory space or (ii) have low efficiency.
However, the magnitude of real-world KGs is much larger.
For instance, a real-world KG YAGO3 includes $\sim$17 million entities, while one of the largest existing benchmarks IDS100K~\cite{OpenEA2020VLDB} only extracts $\sim$0.6\% entities from YAGO3.


To scale up the EA methods for large-scale KGs,
a prevalent approach is to train them on a cluster of machines.
Nonetheless, \emph{the cost of training an EA model on a cluster of machines is prohibitory.}
First, it is unaffordable for many users to purchase a cluster of machines.
Second, it is challenging for ordinary users to deal with cluster management and unpredictable emergencies~\cite{GraphChi12}.
Third, distributed EA necessitates collecting the training results from different machines, and such overhead is not negligible.

The obstruction with distributed computing provokes us to partition an EA dataset into multiple mini-batches and train the samples in each mini-batch independently.
It greatly saves the hardware cost as a stand-alone machine equipped with a GPU is able to run the EA model when the input dataset (i.e., a mini-batch) is of small or moderate size. Furthermore, it requires \emph{zero} coordination among multiple machines since all the training results are stored locally.
Despite these benefits, aligning entities for large-scale KGs in a mini-batch fashion, however, is still a challenging endeavor.

\vspace{0.05in}
\noindent
\textbf{Challenge \uppercase\expandafter{\romannumeral1}:} \emph{How to effectively generate mini-batches?}
A straightforward method is to partition the entire dataset into several random subsets.
Because of its simplicity, it is a common practice used in various tasks, e.g., word translation~\cite{CSLS} and text classification~\cite{fasttext_classification}.
In those tasks, a dataset can be randomly divided into several mini-batches due to the mutual independence between data.
On the contrary, EA is highly related to the structures of KGs.
Random partition destroys KG's structure and thus adversely affects the EA results, as verified in the experiments to be presented in Section~\ref{sec:exp_minibatchgeneration}.
Apart from the importance of maintaining the structure of each KG when generating mini-batches, it is equally crucial to allocate the possibly equivalent entities to the same mini-batch.
If two equivalent entities are placed into different mini-batches, they cannot be aligned.
However, maintaining the graph structure and meanwhile allocating the potentially equivalent entities to the same mini-batch is challenging, as demonstrated by Example~\ref{example_intro2}.

\begin{example}
\label{example_intro2}
\vspace{-1.5mm}
Figure \ref{fig:example} depicts two KGs (i.e., $KG_{EN}$ and $KG_{FR}$) containing equivalent entities.
Each KG is divided into two mini-batches.
Entities highlighted in the same color should be assigned to the same mini-batch.
$KG_{EN}$ and $KG_{FR}$ are heterogeneous.
It is worth noting that there is a wide range of highly heterogeneous KGs in real-life.
Two partition strategies are used to generate mini-batches, represented by red dotted lines and blue dotted lines respectively.
The \emph{red dotted line} indicates that each KG is divided by minimizing edge-cut to minimize the structural loss.
Due to the graph heterogeneity, some equivalent entities are assigned to different mini-batch in this case, such as ``T-Minus (producer)'' of $KG_{EN}$ and ``T-Minus'' of $KG_{FR}$.
The \emph{blue dotted line} symbolizes that each KG is divided by preserving the equivalent entities into the same mini-batch.
Nonetheless, many edges are cut, leading to the loss of structural features and poor EA results.
\end{example}







\vspace*{-1.5mm}
\noindent
\textbf{Challenge \uppercase\expandafter{\romannumeral2}:} \emph{How to recoup the loss of accuracy inevitably caused by the mini-batch generation?}
Since different real-world KGs are heterogeneous, even a perfectly designed mini-batch generation method will inevitably lose seeds or destroy KG's structures, thereby reducing the EA performance, as mentioned in Example \ref{example_intro2}.
Meanwhile, it is widely acknowledged that the \emph{name information} of entities undoubtedly improves the EA performance~\cite{DegreeAware20, AttrGNN20, BERT-INT20}.
Also, \cite{tkdeSurvey20} has stated that several real-life entities are difficult to be aligned based purely on their structural features but are easy to be matched by their name information.
In addition, it is common that real-life entities from different KGs (e.g., YAGO and DBpedia) share the same naming convention, which further makes it practical to utilize entities' name information for EA.
The power of name information inspires us to explore that whether the use of the entity name could complement the seeds or the alignment features for mini-batch training.
The existing name-based EA methods \cite{BERT19, DegreeAware20, AttrGNN20, HGCN19} tend to use a pre-trained language model (e.g., BERT~\cite{BERT19}) to initialize entity embeddings with their name features and then fine-tune these informative embeddings.
As mentioned before, training an EA model with large-scale KGs is challenging, making it impractical to fine-tune the name-based entity embeddings for large-scale EA.
Accordingly, we are required to exploit the name features to facilitate large-scale EA in an efficient and lightweight way.



To address these challenges, we propose \textsf{LargeEA} to align entities between large-scale KGs.
\textsf{LargeEA} consists of two pivotal channels:
(i) \emph{structure channel}, which is introduced to learn the structural features of entities in a mini-batch fashion; and
(ii) \emph{name channel}, which is presented to efficiently
augment the alignment results based on the entities' name features.
Thereafter, \textsf{LargeEA} fuses both the structural features and the name features to produce the final EA results.
Our contributions are summarized as follows:

\begin{figure}[t]
\centering
\includegraphics[width=3.3in]{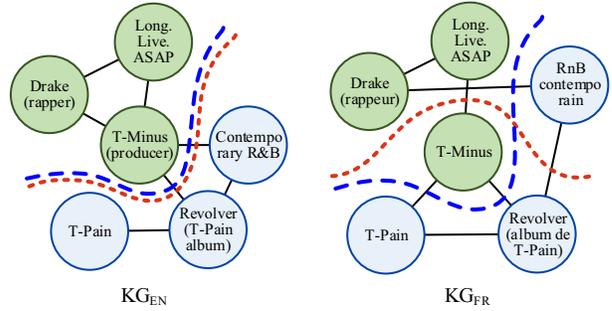}
\vspace{-4mm}
\caption{Example of mini-batch generation for EA}
\label{fig:example}
\vspace{-6mm}
\end{figure}

\begin{itemize}
    \item{\emph{Large-scale EA framework.}}
    \textsf{LargeEA}~\cite{LargeEA} is the first EA framework that aligns entities between large-scale KGs by fusing features from \emph{structure channel} and \emph{name channel}.
    Any EA model suffering from scalability issue can be easily integrated into \textsf{LargeEA} to deal with large-scale EA (Section~\ref{sec:overview}).
    \item{\emph{Memory saving EA channels.}}
    In the \emph{structure channel}, we propose a memory saving \emph{METIS-CPS} strategy to support mini-batch training under the premise of minimizing the loss of structural features and seed alignment (Section~\ref{sec:structure_channel}).
    In the \emph{name channel}, we introduce \emph{NFF} and a \emph{name-based data augmentation} to
    make effective use of name features without any complex training process (Section~\ref{sec:name_channel}).
    \item{\emph{Large-scale EA benchmark.}}
    Since no public large-scale EA benchmark is available, we develop DBP1M~\cite{LargeEA},
    a large-scale EA benchmark extracted from real-world KGs (Section~\ref{sec:exp}).
    \item{\emph{Extensive experiments.}} We conduct comprehensive experimental evaluation on EA tasks against state-of-the-art approaches over both the existing EA benchmarks and newly proposed DBP1M. Extensive experimental results demonstrate the superiority of \textsf{LargeEA} (Section~\ref{sec:exp}).
\end{itemize}




\section{Our Framework}
\label{sec:framework}



\subsection{Problem Statement and LargeEA Overview}\label{sec:overview}
A knowledge graph (KG) can be denoted as $G = (E,R,T)$, where $E$ is the set of entities, $R$ is the set of relations, and $T=\{(h,r,t)~|~h,t \in E, r \in R\}$ is the set of triples, each of which represents an edge flowing from an entity $h$ to another entity $t$ via a relation $r$.
Entity alignment (EA) \cite{OpenEA2020VLDB} aims to find the 1-to-1 mapping of entities $\psi$ from a source KG $G_s = (E_s,R_s,T_s)$ to a target KG $G_t = (E_t,R_t,T_t)$.
Formally, $\psi = \{(e_s, e_t) \in E_s \times E_t~|~e_s \equiv e_t\}$, where
$e_s \in E_s$, $e_t \in E_t$, and $\equiv$ means an equivalence relation between $e_s$ and $e_t$.
In most cases, a small set of equivalent entities $\psi^{\prime} \subset \psi$ is known beforehand and can be used as seed alignment (training data).
A representative experimental study \cite{OpenEA2020VLDB} has indicated that, using a small set of seed alignment (e.g., 20\% of the total number of aligned entities) as training data conforms to the real-world.

\begin{figure}[t]
\centering
\hspace{-2mm}
\includegraphics[width=3.5in]{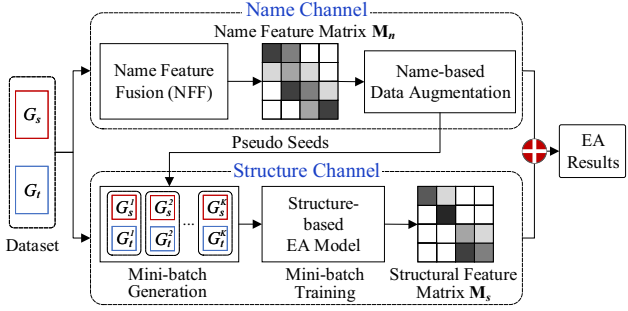}
\vspace{-7mm}
\caption{LargeEA framework}
\label{fig:framework}
\vspace*{-7mm}
\end{figure}

We summarize the main components of the framework \textsf{LargeEA} in Figure~\ref{fig:framework} to provide an overview.
\textsf{LargeEA} takes as inputs a source KG $G_s$ and a target KG $G_t$, and performs EA with the help of the \emph{structure channel} and the \emph{name channel}.
In the structure channel, both $G_s$ and $G_t$ are divided into $K$ subgraphs via the proposed mini-batch generation method METIS-CPS.
It is designed to increase entity locality so that most entities can find their equivalence within the same mini-batch.
After generating mini-batches, we use an EA model to learn the structural similarities between entities in each mini-batch locally.
In the name channel, \textsf{LargeEA} presents
a name feature fusion approach (NFF) to evaluate the name similarity between entities.
Besides, we use a simple but highly effective \emph{data augmentation} to generate pseudo seeds, which complements the loss of seeds caused by the mini-batch generation process in the structure channel.
Finally, we further fuse the structural similarity and the name similarity between entities to derive the EA results. 
%

\begin{figure*}[t]
\centering
\includegraphics[width=7in]{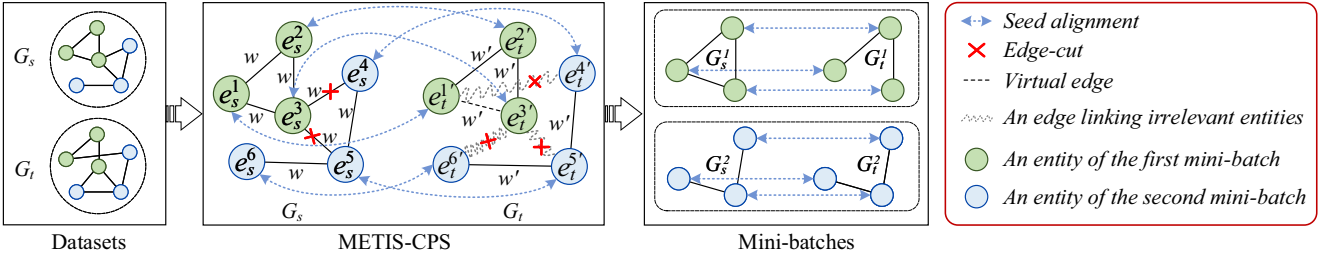}
\vspace*{-2mm}
\caption{A toy example of METIS-CPS workflow. For simplicity, we only depict how entities contained in the seed alignment are partitioned, although all entities in the two input KGs are actually involved in this workflow.}
\label{fig:metis-cps}
\end{figure*}

\subsection{Structure Channel}\label{sec:structure_channel}


\subsubsection{Mini-batch Generation}\label{sec:mini-batch-generation}
Training an EA model in a mini-batch fashion is associated with the following two conditions:
(i) partitioning a large-scale KG into $K$ subgraphs;
and (ii) placing every entity and its potential equivalence in the same mini-batch.
To generate mini-batches under these conditions, we introduce two strategies, i.e., the \uline{v}anilla \uline{p}artition \uline{s}trategy (VPS) and the \uline{METIS}-based \uline{c}ollaborative \uline{p}artition \uline{s}trategy (METIS-CPS).

\noindent
\textbf{VPS.}
It first allocates the seed alignment $\psi^{\prime}$ into $K$ mini-batches equally and then randomly adds the remaining entities from $G_s$ and $G_t$ into the $K$ mini-batches.
The time and space complexities of VPS are both $O(|E_s| + |E_t|)$.
VPS ensures that each mini-batch contains the same number of seeds, as an imbalanced distribution of seeds leads to poor training performance of the EA model in some mini-batches, e.g., it is unable to train the EA model without any seed alignment in a mini-batch.

\noindent
\textbf{METIS-CPS.}
Although simple, VPS relies on random partitioning, which may destroy the structure of each KG.
To this end, we present METIS-CPS. The workflow is shown in Figure~\ref{fig:metis-cps}.

We first review the main idea of METIS~\cite{METIS98}, from which our strategy is derived.
METIS aims to divide a graph into $K$ subgraphs that obey the \emph{modularity maximization principle}.
Following this principle, METIS guarantees that the sum of the weights of the edge-cuts is minimized.
Thus, each entity and its neighbors can be clustered into the same partition to a large extent, while irrelevant entities are located in different partitions.
For simplicity, we denote the weight of an edge between entities $e_i$ and $e_j$ as $w(e_i, e_j)$, and suppose all the edges in a KG share an equal weight.
In the current implementation, we set every $w(e_i, e_j) = 1$.
The time complexity and space complexity of METIS are $O(|E| + |T| + Klog(K))$ and $O(|E| + |T|)$, respectively~\cite{scalaGWL19}.

We then detail the partition process of METIS-CPS in the following.
First, we deploy METIS~\cite{METIS98} to split $G_s$ into $K$ subgraphs $G_s^{i}$, $i \in \{1,2,...,K\}$.
We denote $L_{s}^{i}$ the set of entities belonging to the seed alignment $\psi'$ that are contained in $G_s^{i}$.
Take Figure~\ref{fig:metis-cps} as an example.
After performing METIS, $G_s$ is partitioned into 2 subgraphs, i.e., $G_s^{1}$ and $G_s^{2}$. We have $L_{s}^{1} = \{e_s^{1}, e_s^{2}, e_s^{3}\} \in G_s^{1}$ and $L_{s}^{2} = \{e_s^{4}, e_s^{5}, e_s^{6}\} \in G_s^{2}$.
Next, we explain how to partition $G_t$ into $K$ subgraphs $G_t^{i'}$, $i' \in \{1,2,...,K\}$.
Let $L_{t}^{i'}$ be the set of entities in $G_t$ that are equivalent to the entities in $L_{s}^{i}$. It is preferable that entities in $L_{t}^{i'}$ are all included by $G_{t}^{i'}$.
To achieve this goal, we re-assign appropriate weights to the edges in $G_t$ in two phases.

\vspace{0.5mm}
\noindent
\emph{Phase 1: Increasing weight for relevant entities.}
According to the modularity maximization principle, entities connected by edges with high weight in a dense graph are not likely to be partitioned.
As a result, a plausible intuition for preventing entities to be divided into different mini-batches is that, we can generate a \emph{connected graph} $CG^{i}$ with high weighted edges for entities $L_{t}^{i'}$.
Specifically, given a set of entities $L_{t}^{i'}$ whose equivalent entities $L_{s}^{i}$ belong to the same subgraph $G_{s}^{i}$,
we randomly select $q$ entities from $L_{t}^{i'}$, denoted as $\mathcal{Q}=\{e_1,e_2,...,e_{q}\}$, and make sure all those $q$ entities are able to reach all the other entities in ($L_{t}^{i'}-\mathcal{Q}$) by adding a \emph{virtual edge} between each $e_i \in \mathcal{Q}$ and each $e_j \in L_{t}^{i'}-\{e_i\}$ iff there is no edge between them to make the connected graph $CG^i$ much denser. Thereafter, we reset the weight of each edge of $CG^{i}$ to $w' \gg 1$ to prevent the entities of $L_{t}^{i'}$ from being assigned to different mini-batches.
Since the time cost of the mini-batch generation depends on $q$, we set $q=1$ in the implementation to save time.
This is because empirically $q=1$ is able to achieve satisfactory partition results in our experiments.
Note that, the virtual edge is only used to assist the METIS-CPS in graph partition but not to change the graph structure of the original KG.
In Figure~\ref{fig:metis-cps}, since all the equivalent entities of $e_t^{1'}$, $e_t^{2'}$, and $e_t^{3'}$ belong to $L_{s}^{1} \in G_{s}$, we need to put the three entities into the same subgraph to avoid the destroy of seed alignment.
Thus, we add a virtual edge between $e_t^{1'}$ and $e_t^{3'}$ to form a connected graph $CG^{1}$, and re-assign the weight of edges in $CG^{1}$ to $w'$. Therefore, $e_t^{1'}$, $e_t^{2'}$, and $e_t^{3'}$ are unlikely to be partitioned into different subgraphs.
The time complexity of this phase is $O(|\psi'|+\frac{1}{K} \times |\psi'|^2)$.

\vspace{0.5mm}
\noindent
\emph{Phase 2: Reducing weight for irrelevant entities.}
Let $(e_s^{i}, e_t^{i'})$ and $(e_s^{j}, e_t^{j'})$ denote two seed alignments, respectively. Here, $e_s^{i}, e_s^{j}$ $\in E_s$ and $e_t^{i'}, e_t^{j'} \in E_t$.
Assume that $e_s^{i}$ and $e_s^{j}$ are located in different subgraphs after partitioning, and there is an edge between $e_t^{i'}$ and $e_t^{j'}$. It is possible that the entities $e_t^{i'}$ and $e_t^{j'}$ are assigned into the same subgraph by graph partitioning.
Accordingly, it is required to prevent them from being gathered into the same subgraph to guarantee that those seed alignments are well preserved even after partitioning.
To achieve this purpose, a simple but effective method is to assign zero weight to the edge, i.e., $w(e_t^{i'}, e_t^{j'})=0$.
In Figure~\ref{fig:metis-cps}, we are required to assign $w(e_t^{1'}, e_t^{4'})=0$, $w(e_t^{3'}, e_t^{5'})=0$, and $w(e_t^{3'}, e_t^{6'})=0$.
After performing this phase, we are ready to divide the target KG $G_t$ into subgraphs by executing the METIS strategy. The time complexity is $O\left(\frac{(K-1)|\psi'|^2}{K^2} + |E_t|+|T_t|+Klog(K)\right)$.

Finally, each mini-batch can be generated by putting together a subgraph of $G_s$ and another subgraph of $G_t$ that contain the most number of seed alignments.
The total time complexity and space complexity of METIS-CPS are $O\large(|\psi'|+\frac{(2K-1)|\psi'|^2}{K^2} + |E_s|+|E_t|$ $+|T_s|+|T_t|+Klog(K)\large)$ and $O(|E_s|+|E_t|+|T_s|+|T_t|)$, respectively.



\vspace{-2mm}
\subsubsection{Mini-batch Training}\label{sec:mini-batch-training}
\textsf{LargeEA} treats mini-batch training as a black box and users have the flexibility to utilize any existing EA model to learn a set of embeddings that can be used to represent the structural features of entities.
For the EA task, many GNN-based methods \cite{AttrGNN20,KECG19, EVA20, AliNet20} have achieved promising performance by propagating the alignment signal to the entity’s neighbors.
Inspired by this, we propose to incorporate GNN-based models into \textsf{LargeEA}.
Our current implementation includes two representative GNN-based EA models, i.e., GCN-Align~\cite{GCN-Align18} and RREA~\cite{RREA20}.

For ease of understanding, we sketch the core idea of how \textsf{LargeEA} employs the existing GNN-based EA models.
Generally, the GNN-based models generate each entity's embedding as follows~\cite{RREA20, GCN17, GAT18}.
Formally, $\bm{h}_{\mathcal{N}_{e_{i}}^{e}}^{l-1} = f(\{\bm{h}_{e_{k}}^{l-1}, \forall e_{k} \in\left\{e_{i}\right\} \cup \mathcal{N}_{e_{i}}^{e}\})$; $\bm{h}_{e_{i}}^{l} = \sigma(\bm{W}^{l-1} \cdot \bm{h}_{\mathcal{N}_{e_{i}}^{e}}^{l-1})$.
Here, $f(\cdot)$ is to aggregate information from the neighbors of every entity;
$\mathcal{N}_{e_{i}}^{e}$ represents the set of neighboring entities around $e_i$;
$\boldsymbol{W}^{l-1}$ is the transformation matrix of layer $l-1$; and $\boldsymbol{h}_{e_{i}}^{l}$ denotes the embedding of $e_{i}$ after performing $l$-layer GNNs.

To maximize the similarities of equivalent entities in each mini-batch, GNN-based EA models often use \emph{triplet loss}. Formally, $\mathcal{L}=\sum\nolimits_{(e_s^{i}, e_t^{i'}) \in \psi'} \left[f_p(\bm{h}_{e_s^{i}},\bm{h}_{e_{t}^{i'}}) + \gamma - f_n(\bm{h}_{e_s^{i}},\bm{h}_{e_{t}^{i'}}) \right]_{+}$.
Here,
$\bm{h}_{e_s^{i}}$ and $\bm{h}_{e_{t}^{i'}}$ represent the embeddings of $e_s^{i}$ and $e_t^{i'}$ learned by a structure-based EA model, respectively;
$f_p(\cdot,\cdot)$ represents the distance between $\bm{h}_{{e}_{s}^{i}}$ and $\bm{h}_{{e}_{t}^{i'}}$;
$f_n(\cdot,\cdot)$ denotes the distance of a negative entity pair derived from $e_s^{i}$ and $e_t^{i'}$, generated by replacing either $\bm{h}_{{e}_{s}^{i}}$ or $\bm{h}_{{e}_{t}^{i'}}$ with a new embedding according to the nearest neighbor sampling~\cite{RREA20}; $[x]_{+} = max\{0,x\}$; and
$\gamma > 0$ is a margin hyper-parameter.

We denote $\bm{M_s}$ the total structure-based entity similarity matrix, where each value is computed by the Manhattan distance.
%
$\bm{M_s}$ is highly sparse.
With independent mini-batch training, all non-zero similarity values lie on the diagonal blocks of $\bm{M_s}$.
It saves memory cost for coping with large-scale EA. The memory cost of storing $\bm{M_s}$ is $\mathcal{O}(|E_s|)$, i.e., the number of entities in the source KG.
The time and space complexities of the entire mini-batch training process are $O\left (|\psi'|\times(|T_s|+|T_t|)\right)$ and $O\left(|D_{str}|\times(|E_s| + |E_t|) + |T_s| + |T_t|\right)$, respectively.
Here, $|D_{str}|$ denotes the dimension of every entity embedding learned by the mini-batch training.

\subsection{Name Channel}\label{sec:name_channel}

In this section, we first present a name feature fusion method; we then introduce a name-based data augmentation; and we finally describe how to fuse name channel and structure channel for EA.

\noindent
\textbf{Name Feature Fusion.}
As performing the independent training within each mini-batch inevitably causes the loss of certain structural features of KGs, it is essential to incorporate other procedures to complement the loss caused by graph partitioning. In this work, we propose to consider entities' name features as an effective approach, namely NFF, to improve the EA performance.
Given two entity sets $E_s \in G_s$ and $E_t \in G_t$,
NFF computes the name similarities between $E_s$ and $E_t$ by fusing the name features from both \emph{semantic aspect} $\bm{M_{se}}$ and \emph{string aspect} $\bm{M_{st}}$.
Mathematically, $\bm{M_n}$ ($= \bm{M_{se}} + \gamma \bm{M_{st}}$) represents the fused matrix, where $\gamma \in (0,1]$ is a hyper-parameter controlling the contribution of the string-based name similarity to $\bm{M_n}$. In the current implementation, we set $\gamma = 0.05$ since many studies~\cite{ditto20, DeepMatcher18} have argued that semantic feature is much more important than string-based feature.
In the following, we detail SENS and STNS, the two functions to get name similarity according to the above two aspects, respectively.

Function SENS is to get semantic name similarity.
Concretely, we use BERT~\cite{BERTurl} to transform each entity name into a sequence of tokens.
Thereafter, the semantic embedding of each entity can be generated by applying \emph{max-pooling}, which assigns each token an embedding with fixed-dimension for each entity and then picks the maximum value in each dimension among all embedded tokens (related to the entity) to form a new embedding representing the entity.
Let $\bm{S_s}$/$\bm{S_t}$ denote the semantic embedding matrix of $E_s$/$E_t$.
Each embedding $\bm{h}_{e}$
can be normalized by the equation $\bm{h}_{e} = \frac{\bm{h}_{e}}{\left\| \bm{h}_{e} \right\|_2 + \epsilon}$, where $\bm{h}_{e}$ is an entity embedding from $\bm{S_s}$ or $\bm{S_t}$, and $\epsilon > 0$ is to prevent the denominator from being zero.
Since the embeddings of name features are mutually independent, we can randomly split the semantic embedding matrix ($\bm{S_s}$ or $\bm{S_t}$) into $K$ segments for saving memory.
Then, we iteratively find the top-$k$ semantic similar entity pairs (denoted as $\bm{M_{se}^{ij}}$) between any two segments $\bm{S_{s}^{i}}$ and $\bm{S_{t}^{j}}$, where $i,j \in \{1,2,...,K\}$, by Faiss~\cite{JDH17}, an efficient similarity search method with GPU(s) that can cope with large-scale data.
Here, we use Manhattan distance to measure the semantic similarity.
Finally, the complete semantic matrix $\bm{M_{se}}$ can be obtained by only collecting the semantic similarity results computed by Faiss.
The time and space complexities of SENS are $O(|D_{se}|\times|E_s| \times |E_t|)$ and $O\left(|D_{se}|\times(|E_s|+|E_t|)\right)$, respectively.
Here, $|D_{se}|$ denotes the dimension of every entity embedding obtained from BERT. 
Though the time complexity seems high, the GPU-based Faiss greatly speeds up the similarity computation process.

We would like to highlight that it is essential to consider the top-$k$, but not all, pairs of entities with high similarity scores for the EA task. Specifically, entity pairs with low similarity scores are probably the erroneous alignment and they are \emph{not} expected to provide any useful information for the name-based EA task.
In addition, filtering out the low scores of entity pairs but retaining the top-$k$ similarity scores in the similarity matrix $\bm{M_{se}}$ notably reduces the memory cost from $\mathcal{O}(|E_s||E_t|)$ to $\mathcal{O}(k|E_s|)$, with $k \ll |E_t|$.
Here, $\mathcal{O}(|E_s||E_t|)$ refers to the cost of storing the similarity scores of all the candidate entity pairs (in total $|E_s||E_t|$), which is clearly impractical for large-scale KGs.

Function STNS is to measure the string-based name similarity (i.e, Levenshtein distance in the implementation) between entities.
It is well-known that calculating Levenshtein distance for large-scale entity pairs is time-consuming and computationally expensive.
Given two sets of entities $E_s$ and $E_t$,
the time and space complexities of Levenshtein-distance-based similarity computation are both $O\left(|E_s| \times |E_t| \times Max(len_{E_s}) \times Max(len_{E_t})\right)$ in general.
Here, $Max(len_{E_s})$ (w.r.t. $Max(len_{E_t})$) denotes the maximum length of an entity's name from the set $E_s$ (w.r.t. $E_t$).
Our solution is to filter out the pairs that are extremely different, as entity pairs with different names are less likely to be aligned.
Motivated by the efficiency of the datasketch library~\cite{datasketch}
for finding similar entities per entity, we deploy the datasketch library for this purpose.
Concretely, the datasketch library employs MinHash-LSH to reduce the computational cost of finding similar entity pairs.
The time complexity of datasketch is $O(|E_s|)$.
We only retain the entity pairs $P$ whose Jaccard similarities are above $\theta$, the lower bound of the string difference.
Then, we compute the string similarity for each entity pair in $P$ by Levenshtein distance. We denote the string-based similarity matrix as $\bm{M_{st}}$.
Similarly, the benefits of using the threshold $\theta$ are to (i) save the memory space of storing $\bm{M_{st}}$ and (ii) reduce the total number of entity pairs that require Levenshtein distance computation.




\noindent
\textbf{Name-based Data Augmentation.}
Recall that some seed alignments in the training data may be missing after mini-batch generation.
To this end, we describe how we apply \emph{data augmentation} (DA) based on the similarity between entities' name features to
generate seed alignment automatically.
We are inspired by the idea of \emph{cycle consistency} from word translation~\cite{cycleconsistency1970}, which states that, if two sentences from different languages can be translated to each other, they have the same meaning.
Therefore, we generate pseudo seed alignment in accordance with the constraints that two entities are mutually the most similar to each other.

\noindent
\textbf{Channel Fusion for Aligning Entities.}
As highlighted by massive prior studies~\cite{tkdeSurvey20, DegreeAware20, AttrGNN20, CEAFF20}, the name feature and structural feature can complement each other in the task of EA.
Motivated by this, we fuse the name similarity matrix $\bm{M_s}$ and the structure similarity $\bm{M_n}$ derived from name channel and structure channel, respectively.
To balance the importance of both structure channel and name channel, we combine $\bm{M_s}$ and $\bm{M_n}$ with equal weights, and derive the final similarity matrix between $E_s$ and $E_t$, denoted as $\bm{M}$.
Formally, $\bm{M} = \bm{M_s} + \bm{M_n}$.
Though simple and intuitive, this approach effectively fuses the features from both name and structure aspects.
We will demonstrate the effectiveness of channel fusion via experimental study to be presented in Section~\ref{sec:ablation}.

\begin{table}[t]\small
\caption{Statistics of the datasets used in experiments}
\vspace{-2mm}
\label{tb:dataset}
\setlength{\tabcolsep}{1mm}{
\begin{tabular}{ll|l|l|l}
\toprule
\multicolumn{2}{c|}{Datasets}                          & \#Entities          & \#Relations & \#Triples           \\ \hline
\multicolumn{1}{l|}{\multirow{2}{*}{IDS15K}}  & EN-FR & 15,000-15,000       & 267-210     & 47,334-40,864       \\
\multicolumn{1}{l|}{}                         & EN-DE & 15,000-15,000       & 215-131     & 47,676-50,419       \\ \hline
\multicolumn{1}{l|}{\multirow{2}{*}{IDS100K}} & EN-FR & 100,000-100,000     & 400-300     & 309,607-258,285     \\
\multicolumn{1}{l|}{}                         & EN-DE & 100,000-100,000     & 381-196     & 335,359-336,240     \\ \hline
\multicolumn{1}{l|}{\multirow{2}{*}{DBP1M}}   & EN-FR & 1,877,793-1,365,118 & 603-380     & 7,031,172-2,997,457 \\
\multicolumn{1}{l|}{}                         & EN-DE & 1,625,999-1,112,970 & 597-241     & 6,213,639-1,994,876 \\ \bottomrule
\end{tabular}}
\vspace{-5mm}
\end{table}

\section{Experiments}
\label{sec:exp}

In this section, we conduct extensive experiments to verify the effectiveness and efficiency of \textsf{LargeEA}, using six datasets, i.e., (i) four small-scale datasets provided by the state-of-the-art benchmark IDS~\cite{OpenEA2020VLDB}; and (ii) two large-scale datasets generated by us.

\vspace{-4mm}
\subsection{Experimental Settings}\label{sec:exp_setting}

\begin{table*}[t]\small
\caption{Overall EA results on IDS15K and IDS100K
}\label{exp:overall_15K_100K}
\vspace*{-4mm}
\begin{threeparttable}
\setlength{\tabcolsep}{0.6mm}{
\begin{tabular}{l|ccccc|ccccc|ccccc|ccccc}
\toprule
\multicolumn{1}{c|}{\multirow{2}{*}{Methods}} & \multicolumn{5}{c|}{IDS15K$_{EN-FR}$} & \multicolumn{5}{c|}{IDS15K$_{EN-DE}$} & \multicolumn{5}{c|}{IDS100K$_{EN-FR}$} & \multicolumn{5}{c}{IDS100K$_{EN-DE}$} \\ \cline{2-21}
 &  H@1 &  H@5 & MRR & Time &  Mem. & H@1 &  H@5 & MRR & Time &  Mem. &  H@1 &  H@5 & MRR & Time  &  Mem. &  H@1 &  H@5 & MRR & Time  &  Mem.\\ \hline
GCNAlign   & 33.8 & 58.9 & 0.45 & 20.29 & 0.13G & 48.1 & 67.9 & 0.57 & 21.60 & 0.13G & 23.0 & 41.2 & 0.32 & 1225.03 & 1.00G & 31.7 & 48.5 & 0.40 & 1639.49 & 1.00G \\
MultiKE     & 74.9 & 81.9 & 0.78 & 290.58 & 6.52G & 75.6 & 80.9 & 0.78 & 350.85 & 10.52G  & 62.9 & 68.0 & 0.66 & 1277.73  & 16.08G & 66.8 & 71.2 & 0.69 & 1765.08 & 16.08G \\
RDGCN       & 75.5 & 85.4 & 0.80 & 554.31  & 8.02G & 83.0 & 89.5 & 0.86 & 739.72 & 8.02G  & 64.0 & 73.2 & 0.68 & 2852.70 & 16.02G  & 72.2 & 79.4 & 0.76 & 3511.14 & 16.02G \\
RREA       & 80.8 & 96.3 & 0.88 & 139.34 & 4.07G  & 85.8 & 96.8 & 0.91 & 137.08 & 4.07G  & --    &  --   &  --   &--  & -- &      --  &   --  &  --   & --  & -- \\
BERT-INT       & 94.2 & 96.4 & 0.95 & 969.61  & 14.07G & 93.5 & 95.0 & 0.94 & 1044.28  & 14.07G & 92.0 &  94.4  & 0.93  & 6991.65 &  14.07G  &     90.8 &  93.3  &  0.92   & 6999.01 &  14.07G \\
\hline
\textbf{\textsf{LargeEA-G}}$_{\rm{EN}\rightarrow \mathbb{L}}$ & 88.4 & 92.2 & 0.90 & 77.00  & 1.54G & 89.2 & 93.4 & 0.91 & 74.81 & 1.54G & 83.9 & 87.5 & 0.86 & 465.10 &  1.74G  & 85.6 & 89.1 & 0.87 & 465.26 & 1.74G \\
\textbf{\textsf{LargeEA-G}}$_{\mathbb{L}\rightarrow \rm{EN}}$ & 89.9 & 92.9 & 0.91 & 75.86 & 1.54G  & 90.8 & 94.2 & 0.92 & 75.89 & 1.54G  & 84.7  & 87.8 &  0.86 & 450.29 & 1.74G   & 85.8 & 89.2 & 0.87 & 452.48 & 1.74G  \\
\textbf{\textsf{LargeEA-R}}$_{\rm{EN}\rightarrow \mathbb{L}}$ & 88.7 & 91.9 & 0.90 & 95.43  & 1.54G  & 89.2 & 94.0 & 0.91 & 96.32  & 1.54G  & 84.4 & 88.0 & 0.86 & 552.84  & 4.04G & 83.4 & 86.7 & 0.85 & 574.00 & 4.04G \\
\textbf{\textsf{LargeEA-R}}$_{\mathbb{L}\rightarrow \rm{EN}}$ & 89.8 & 92.7 & 0.91 & 98.33  & 1.54G  & 91.1 & 94.9 & 0.93 & 96.31  & 1.54G  & 84.3 & 87.5 & 0.86 &  559.51 & 4.04G  & 86.4 & 89.6 & 0.88 & 577.88  & 4.04G \\
\bottomrule
\end{tabular}}
\begin{tablenotes}
\footnotesize
    \item[1] The symbol ``--'' indicates that the EA model is \textbf{NOT} able to perform the EA task by using the GPU in the experimental conditions because of the memory limitation.
    \item[2] The results of all the competitors are obtained by our re-implementation with their publicly available source code.
    \item[3] $\mathbb{L}$ represents the non-English language. For instance, ${\rm{EN}\rightarrow \mathbb{L}}$ denotes that the language of source KG is English and that of target KG is non-English.
\end{tablenotes}
\end{threeparttable}
\vspace{-4mm}
\end{table*}

\noindent
\textbf{Datasets and Evaluation Metrics.}
We conduct experiments on datasets with different scales from two cross-lingual EA benchmarks, i.e., IDS~\cite{OpenEA2020VLDB} and DBP1M.
Table \ref{tb:dataset} lists the detailed statistics.

(i) \emph{IDS.}
Recent work~\cite{OpenEA2020VLDB} indicates that several EA benchmarks (e.g., DBP15K~\cite{JAPE17} and DWY100K~\cite{BootEA18}) contain much more high-degree entities than real-world KGs do.
Consequently, they generate IDS, which contains four cross-lingual datasets, i.e., English and French (IDS15K$_{EN-FR}$ and IDS100K$_{EN-FR}$), and English and German (IDS15K$_{EN-DE}$ and IDS100K$_{EN-DE}$).

(ii) \emph{DBP1M.}
We create two large-scale cross-lingual datasets extracted from a well-known real-world KG, i.e., DBpedia~\cite{DBPedia}.
Concretely,
we retrieve $\sim$1M ground truth of EA by utilizing the inter-language links (ILLs) and \textsf{owl:sameAs} among DBpedia’s multilingual versions, i.e., English and French (DBP1M$_{EN-FR}$), and English and German (DBP1M$_{EN-DE}$).
Unlike the IDS benchmark that ensures the number of entities from one KG is equivalent to that from another KG, we allow KGs to have different number of entities.
For example, the English KGs of our proposed DBP1M benchmark contains more entities.
This conforms to the real-world KGs since the English version of DBpedia is more complete than the versions in other languages.
To simulate the real-world EA scenarios, we also inject \emph{unknown entities}, which cannot find any equivalence based on the EA ground truth, into every dataset of DBP1M.
This is because, it is common that only partial entities can find their equivalence in an EA dataset in real life.
Specifically, we add \emph{unknown entities} that have at least 5 entities (each of which has its equivalent entity in corresponding ground truth datasets) in their neighborhood into each KG, following Sun et al.~\cite{JAPE17}.

We follow \cite{OpenEA2020VLDB} to use 20\% as training data, which conforms to the real world.
The remaining data (80\%) is to test the EA performance.
We use  Hits@$N$ ($N$=1, 5, H@$N$ for short), Mean Reciprocal Rank (MRR), running time (in seconds for small-scale datasets and hours for large-scale ones), and the maximum GPU Memory cost (\emph{Mem.} for short, in GB) as the evaluation metrics.
Here, the running time means the training time of every EA approach.
Higher Hits@N and MRR indicate better performance.
We use the NVIDIA Nsight Systems to monitor the usage of GPU memory.

\noindent
\textbf{Competitors.}
We compare \textsf{LargeEA} with several widely used EA models, which have presented promising EA performance:
(i) \emph{GCNAlign}~\cite{GCN-Align18}, an attribute-powered EA model that uses vanilla GCNs and entity attributes to learn entity embeddings for alignment;
(ii) \emph{MultiKE}~\cite{MultiKE19}, a side-information-based EA model that unifies multiple views of entities;
(iii) \emph{RDGCN}~\cite{RDGCN19}, an EA model that first uses entity names to initialize entity embeddings and then learns these embeddings via a relation-aware dual graph convolutional network;
(iv) \emph{RREA}~\cite{RREA20}, a GNN-based EA model that leverages relational reflection transformation to obtain relation specific embeddings for each entity;
and (v) \emph{BERT-INT}~\cite{BERT-INT20}, which uses BERT~\cite{BERT19} to discover the semantic features contained in the side information of entities instead of considering the graph structure of KGs.

\noindent
\textbf{Variants of \textsf{LargeEA}.}
Note that both GCNAlign and RREA provide variants that purely utilize the structural features to align entities. The former includes a vanilla GCN to learn structural features. The latter provides a GNN-based model that achieves the state-of-the-art EA performance.
Since \textsf{LargeEA} can be easily integrated with structural-based EA models, we present two versions of \textsf{LargeEA}, i.e., \textsf{LargeEA-G} that includes the variant of GCNAlign and \textsf{LagreEA-R} that incorporates the variant of RREA.


\noindent
\textbf{Implementation Details.}
We detail the hyper-parameters used in \textsf{LargeEA} as follows. All the hyper-parameters are set without special instructions.
\emph{In the name channel},
we set the string-based similarity threshold $\theta = 0.5$ and the semantic-based similarity threshold $\phi = 50$ in NFF.
Also, we fix the dimension of every entity embedding obtained by BERT (i.e., $D_{se}$) to be $768$.
\emph{In the structure channel}, we set the number of mini-batches $K = 5$ for IDS15K, $K = 10$ for IDS100K, and $K = 20$ for our DBP1M dataset by default.
Unless explicitly specified, we use RREA as the default EA model in the structure channel and optimize it with Adam for 100 epochs in each mini-batch.
Besides, following the settings in~\cite{GCN-Align18, BERT19}, we set the dimension of every entity embedding generated by the structure channel $|D_{str}| = 200$ for \textsf{LargeEA-G} and $|D_{str}| = 100$ for \textsf{LargeEA-R}.
All experiments were conducted on a personal computer with an Intel Core i9-10900K CPU, an NVIDIA GeForce RTX3090 GPU and 128GB memory. The programs were all implemented in Python.

\subsection{Overall Performance}


\begin{table}[t]\small
\caption{Overall EA results on DBP1M}\label{exp:overall_1M}
\vspace{-4mm}
\setlength{\tabcolsep}{0.1mm}{
\begin{tabular}{l|ccccc|ccccc}
\toprule
\multicolumn{1}{c|}{\multirow{2}{*}{Methods}} & \multicolumn{5}{c|}{DBP1M$_{EN-FR}$}  & \multicolumn{5}{c}{DBP1M$_{EN-DE}$}  \\ \cline{2-11}
\multicolumn{1}{c|}{} & \multicolumn{1}{c}{H@1} & \multicolumn{1}{c}{H@5} & \multicolumn{1}{c}{MRR} & \multicolumn{1}{c}{Time} &  \multicolumn{1}{c|}{Mem.} & \multicolumn{1}{c}{H@1} & \multicolumn{1}{c}{H@5} & \multicolumn{1}{c}{MRR} & \multicolumn{1}{c}{Time} & Mem. \\ \hline
\textbf{\textsf{LargeEA-G}$_{\mathbf{EN}\rightarrow \mathbb{L}}$} & 51.8 & 58.3 & 0.55 & 3.36  & 6.61G  & 55.3 & 60.8 & 0.58 & 2.59 & 4.59G \\
\textbf{\textsf{LargeEA-G}$_{\mathbb{L}\rightarrow \mathbf{EN}}$} & 50.6 & 56.5 & 0.53 & 3.39 & 8.00G & 55.5 & 61.3 & 0.58 & 2.63 & 5.36G \\
\textbf{\textsf{LargeEA-R}$_{\mathbf{EN}\rightarrow \mathbb{L}}$} & 52.8 & 58.7 & 0.56 & 3.58 & 21.15G & 56.1 & 61.3 & 0.59 & 2.88 & 16.01G \\
\textbf{\textsf{LargeEA-R}$_{\mathbb{L}\rightarrow \mathbf{EN}}$} & 51.5 & 57.0 & 0.54 & 3.71 & 21.17G & 56.2 & 61.8 & 0.59 & 2.91 & 16.01G\\
\bottomrule
\end{tabular}}
 \vspace*{-4.8mm}
\end{table}

\subsubsection{Performance on IDS}\label{sec:exp_ids}
Table~\ref{exp:overall_15K_100K} summarizes the EA performance on IDS15K and IDS100K.
We first focus on the \textbf{accuracy evaluation} for the two variants of \textsf{LargeEA} and its competitors.
First, both variants of \textsf{LargeEA} perform better than the existing EA models that also explore both name features and structural features. It validates the superiority of the way how the name feature and structural feature are fused in our framework.
Second, compared to BERT-INT, which achieves state-of-the-art accuracy on the small-scaled datasets (i.e., IDS15K and IDS100K), \textsf{LargeEA} gains up to \textbf{9x GPU memory saving}.
The reason is that BERT-INT highly relies on BERT~\cite{BERT19} that has a much more complex model structure and needs to store more parameters in the training process.
Moreover, BERT-INT suffers from scalability issue due to the models' inherent characteristics.
Different from the other competitors that mainly utilize GPU for model training, the complex model design of BERT forces BERT-INT to store a part of parameters into RAM; otherwise, BERT-INT cannot run successfully because of limited GPU memory.
To be more specific, for IDS15K, BERT-INT requires 14GB GPU memory and 7GB CPU memory on average.
For IDS100K, BERT-INT requires 14GB GPU memory and 58GB CPU memory on average.
Although the usage of GPU memory seems to be stable on datasets with different scales (since we set the same number of mini-batches for BERT-INT on both IDS15K and IDS100K datasets, as suggested by its original paper), the bottleneck of BERT-INT is the tremendous CPU memory requirement.
As expected, it needs at least 580GB RAM for handling a large-scale dataset (e.g., DBP1M) whose size is more than 10 times that of IDS100K.
To further verify the above observation, we have tried to run BERT-INT on DBP1M but failed even with a 128GB of RAM.
Third, we can observe that the influence of the source KG selection on H@1 varies from 0.1\% to 3\%.
Since different KGs are heterogeneous, it is common that selecting different KGs as sources leads to different EA accuracy (up to 4.7\% on some existing models), as confirmed by DGMC~\cite{DGMC20}.


We now turn our attention to the \textbf{running time evaluation}.
We can observe that the training of \textsf{LargeEA} (both variants) is faster than that of other existing EA models.
Particularly, \textsf{LargeEA} runs more than \textbf{10x faster} than BERT-INT, the state-of-the-art EA model.
Recall that BERT-INT is extremely complex in the model structure.
It is time-consuming to obtain reliable parameters in the process of training such a complex model.
This shows the superiority of \textsf{LargeEA} in terms of running time.
One exception is the performance of GCNAlign in IDS15K benchmark. The running time of \textsf{LargeEA-G} is larger than that of the GCNAlign alone.
The reason is that \textsf{LargeEA-G} effectively captures the entity name's features, which requires additional running time.
Despite the fast running time of GCNAlign in the small-scale IDS15K benchmark, \textsf{LargeEA-G} accelerates the running time up to 3.5x than GCNAlign when performing EA in IDS100K, a relatively larger benchmark.

\subsubsection{Performance on DBP1M}
Table~\ref{exp:overall_1M} reports the overall EA performance of \textsf{LargeEA-G} and \textsf{LargeEA-R} on DBP1M.
Note that the EA results produced by \textsf{LargeEA}'s competitors are not reported, as all the competitors fail to perform EA task for any dataset of the large-scale DBP1M in the experiments. The main reason is that they all require a substantial amount of memory space.
In contrast, it is observed that \textsf{LargeEA} is capable of scaling the current EA models to deal with large-scale entity alignment tasks.
As discussed in Section~\ref{sec:mini-batch-training}, the reason is that the mini-batch training of \textsf{LargeEA} greatly saves the memory cost.

\subsubsection{Scalability Evaluation}
\begin{figure}[t]
\centering
\includegraphics[width=0.46\textwidth]{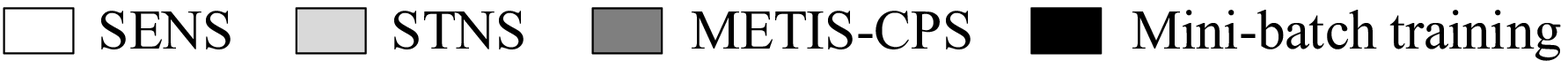}\vspace*{-4.5mm}\\
\subfigure[EN-FR]{
 \includegraphics[width=1.55in]{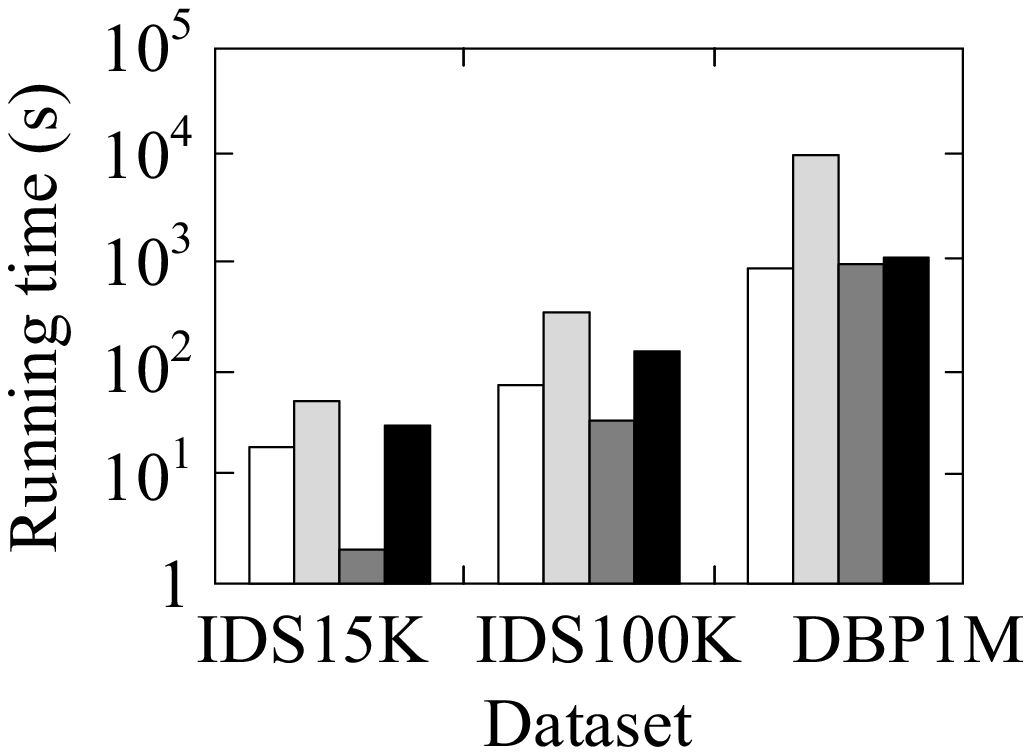}
}
\subfigure[EN-DE]{
 \includegraphics[width=1.55in]{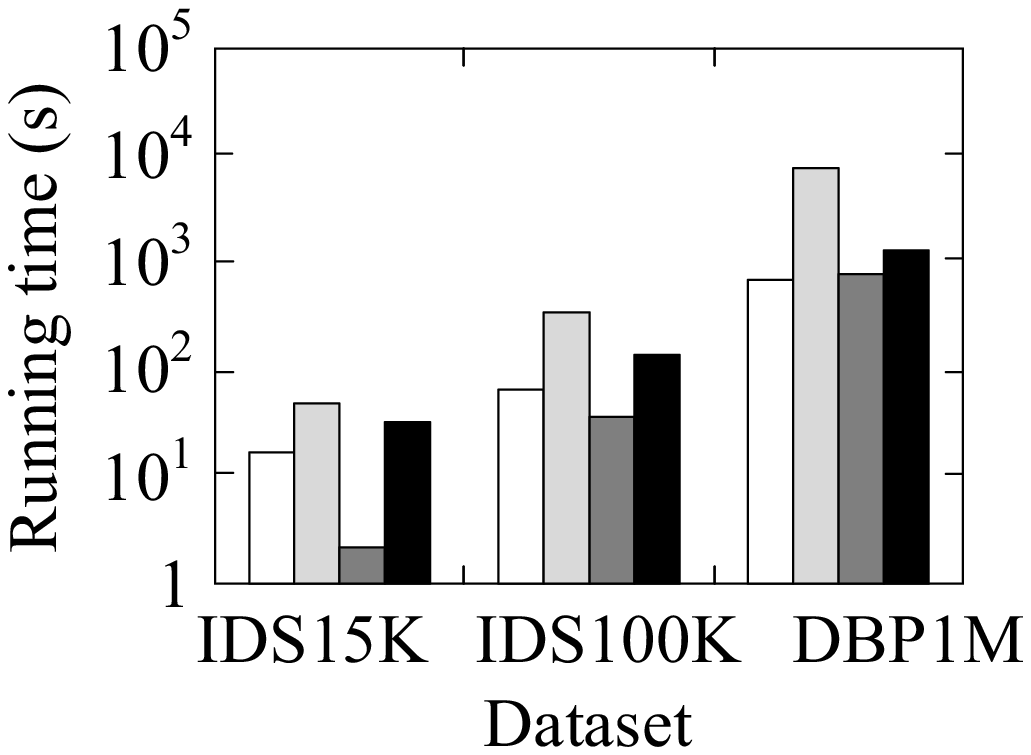}
}
\vspace*{-5mm}
\caption{Scalability analysis vs. datasize}
\vspace*{-3mm}
\label{fig:scalability}
\end{figure}

To further investigate the scalability of our \textsf{LargeEA} framework, we evaluate the running time of each channel of LargeEA on datasets with different scales.
For the structure channel, we report the running time of computing entities' semantic name similarities (SENS) and that of computing entities' string-based name similarities (STNS), respectively.
For the structure channel, we report the running time of mini-batch generation strategy (i.e., METIS-CPS) and that of EA model training.
As depicted in Figure~\ref{fig:scalability}, we observe that the running time of each component increases almost linearly as dataset size grows.
This confirms the scalability of \textsf{LargeEA}.

\begin{figure}[t]
\centering
\includegraphics[width=0.48\textwidth]{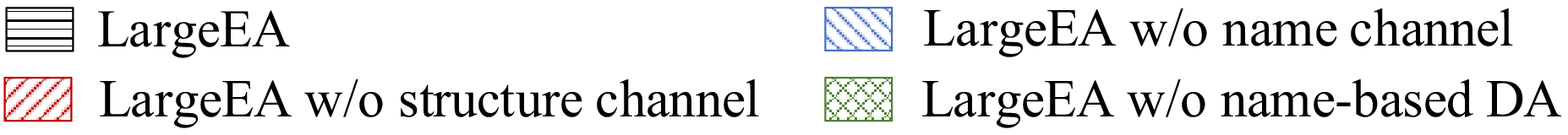}\vspace*{-3.5mm}\\
\subfigure[EN-FR]{
 \includegraphics[width=1.55in]{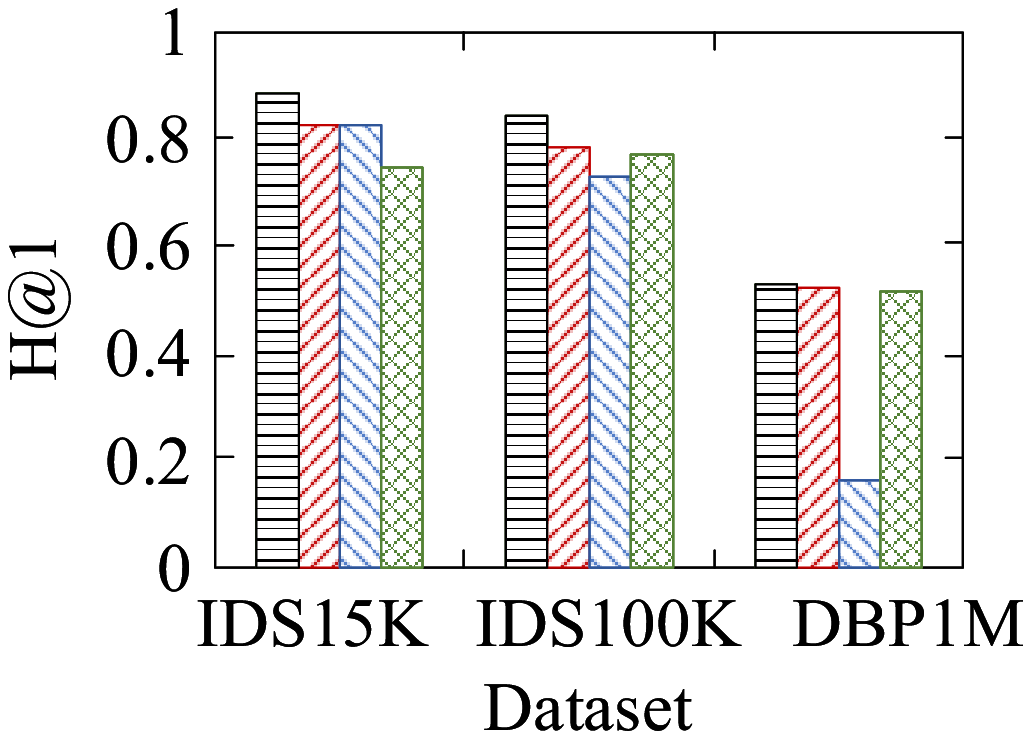}
}
\subfigure[EN-DE]{
 \includegraphics[width=1.55in]{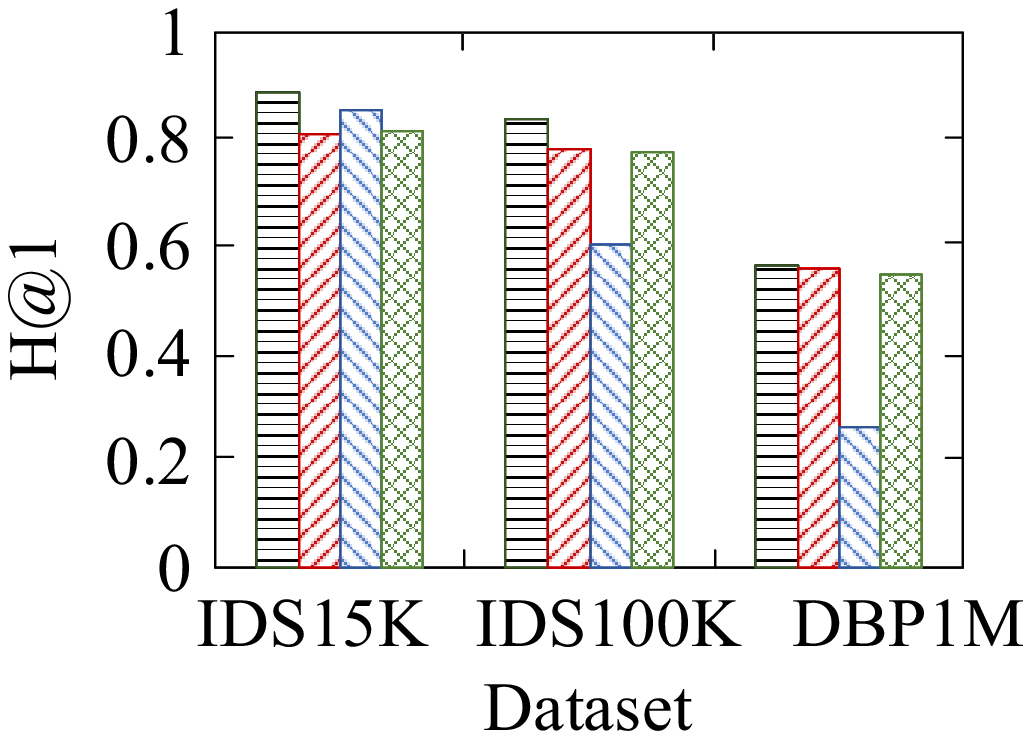}
}
\vspace*{-4mm}
\caption{Ablation studies}
\label{fig:ablation}
 \vspace*{-5mm}
\end{figure}

\subsection{Ablation Study}\label{sec:ablation}
We conduct ablation studies on all datasets,
with results plotted in Figure~\ref{fig:ablation}.
By removing the structure channel, the accuracy of \textsf{LargeEA} drops on $H@1$.
This verifies that the structure channel is an indispensable part of \textsf{LargeEA}.
We also observe that the structural channel has less influence for EA on DBP1M, compared to IDS. It is attributed to the following two reasons.
First, the different number of entities from each side easily leads to more heterogeneous KGs, compared to the IDS where the source KG and the target KG share the same number of entities.
Since the performance of structure-based EA methods highly relies on the graph isomorphism~\cite{DGMC20}, it is more challenging to learn reliable EA signals from heterogeneous KGs, and thus results in relatively worse EA performance.
Second, DBP1M contains \emph{unknown entities}, which further exacerbate the heterogeneity of the source KG and the target KG.
By removing the name channel,
it is observed that the accuracy decline varies from $3\%$ to $37\%$ on different benchmarks.
This verifies that capturing name features of entities and generating pseudo seed alignment greatly improve alignment results.
By removing the name-based data augmentation (DA), the accuracy decline varies from $2\%$ to $14\%$ on different benchmarks.
In particular, the accuracy drops more significantly on IDS15K and IDS100K, compared to that on DBP1M.
This is because IDS15K and IDS100K have richer structural features than DBP1M.
Specifically, the name-based DA provides more seeds for improving the EA results of the structural channel. The richer the structural features, the greater the improvement of EA performance that can be brought by seeds.
Furthermore, EA accuracy on DBP1M is lower than that on IDS15K, attributed to the huge amount of unknown entities.
Normal entities may be aligned to these unknown entities, incurring the drop of $H@1$ when evaluating the results according to the ground truth, which only contains known equivalent entities.

\subsection{Mini-batch Generation Analysis}
\label{sec:exp_minibatchgeneration}

\begin{figure}[t]
\centering
\includegraphics[width=0.46\textwidth]{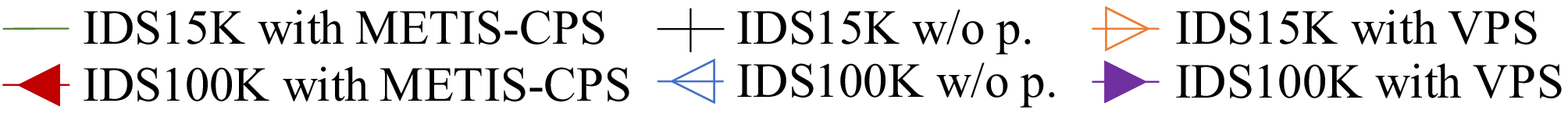}\vspace*{-3mm}
\\
\subfigure[EN-FR]{
    \includegraphics[width=1.54in]{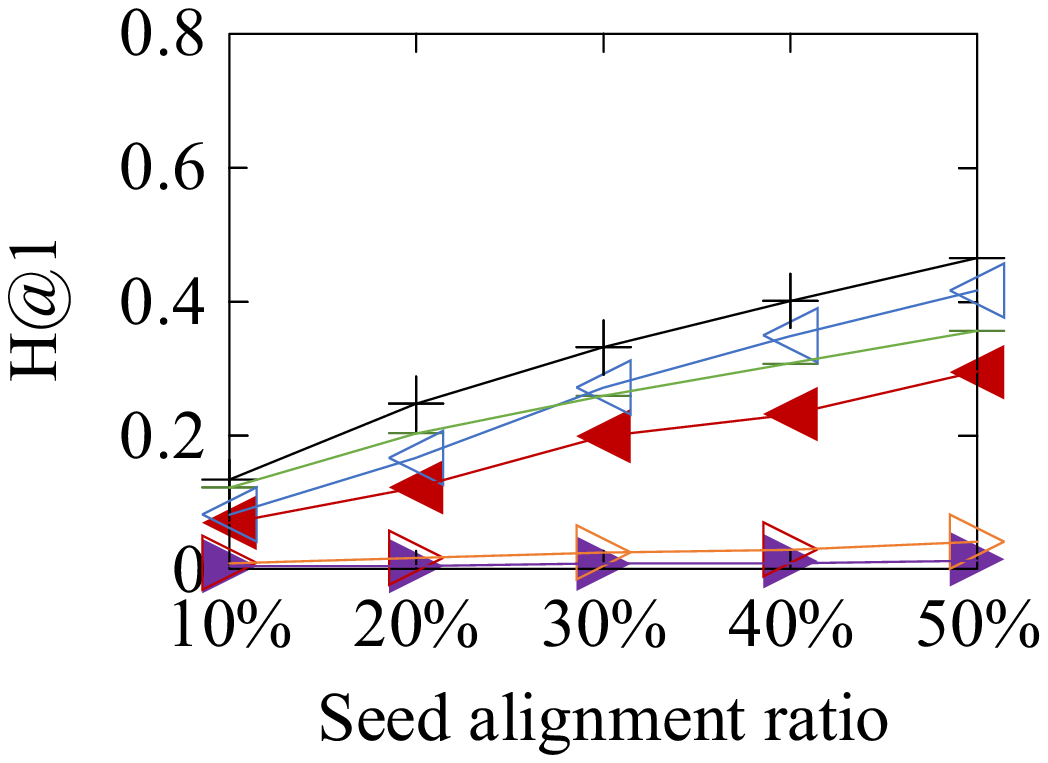}
}\hspace{-2mm}
\subfigure[EN-DE]{
 \includegraphics[width=1.54in]{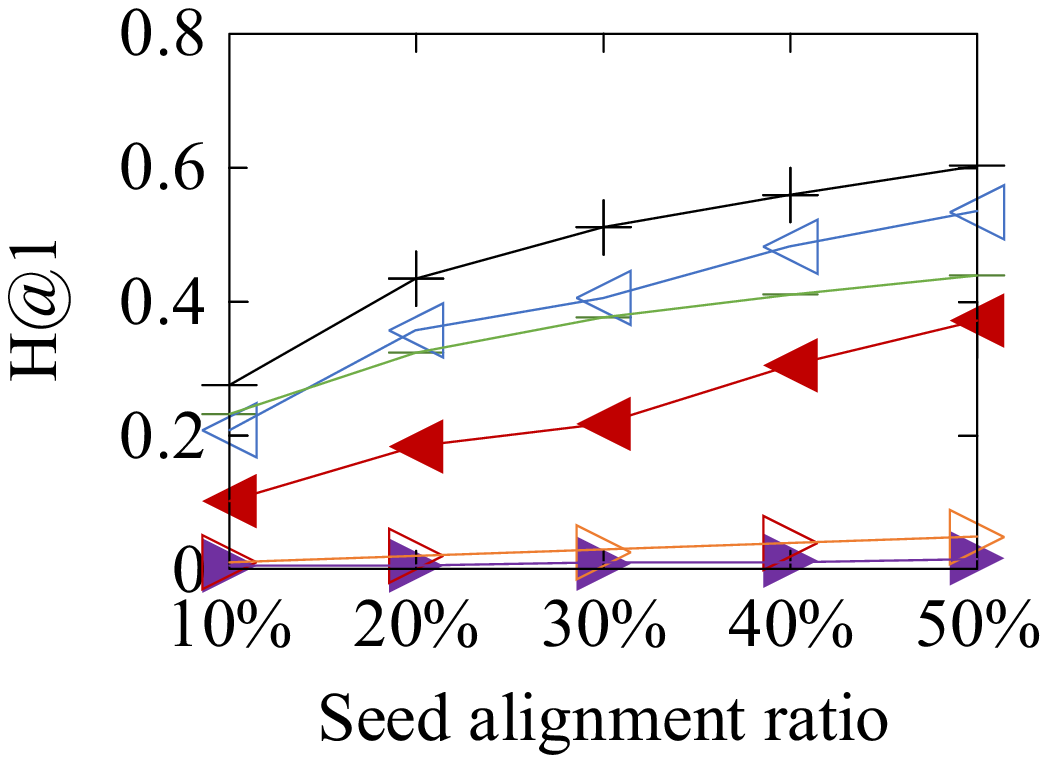}
}\vspace*{-2mm}
\\
\hspace{1mm}
\includegraphics[width=0.46\textwidth]{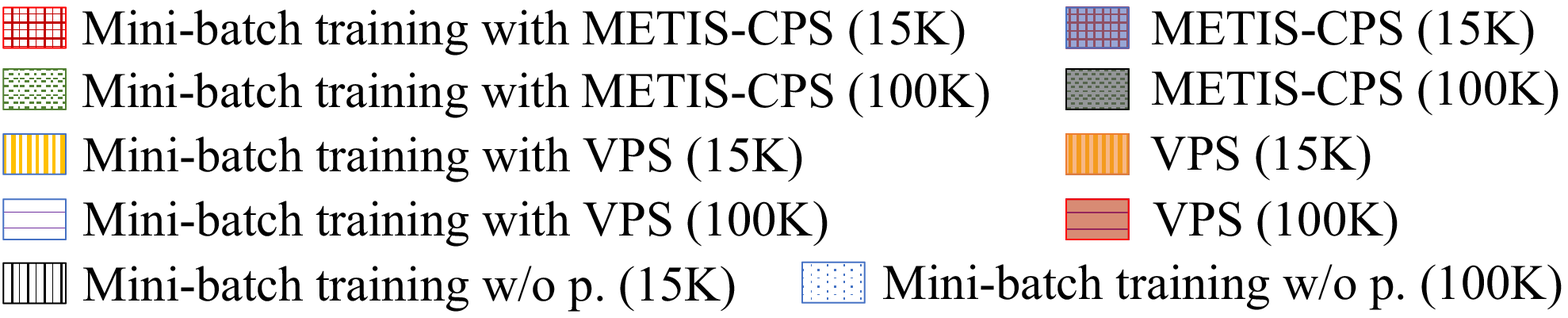}\vspace*{-3mm}
\\
\subfigure[EN-FR]{
    \includegraphics[width=1.54in]{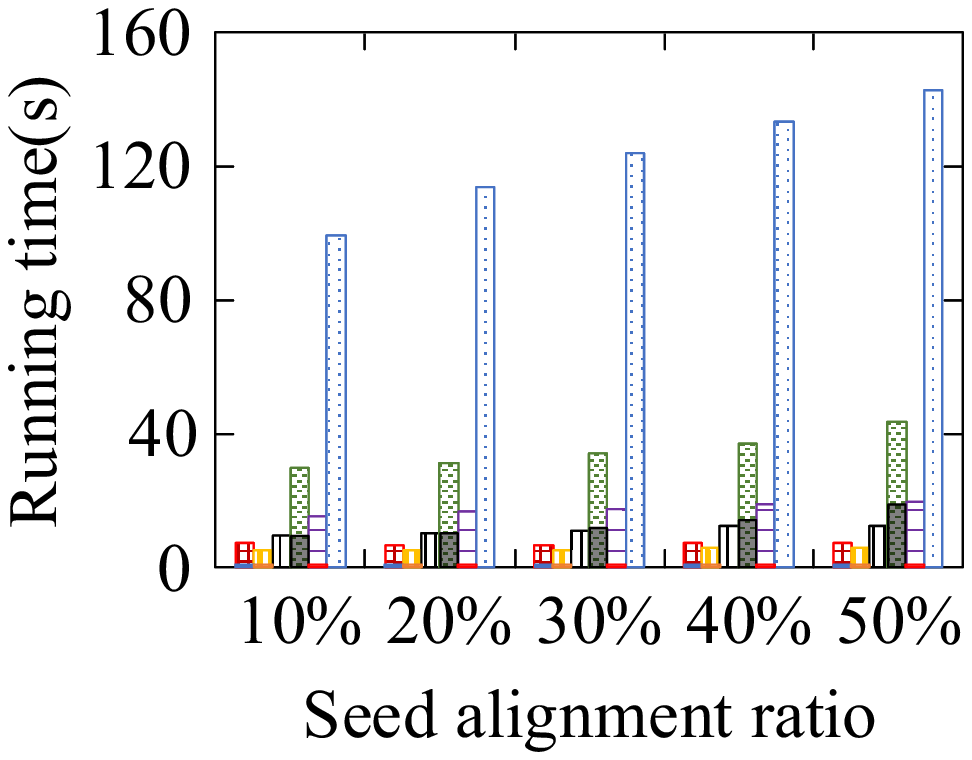}
}
\hspace{-2mm}
\subfigure[EN-DE]{
 \includegraphics[width=1.54in]{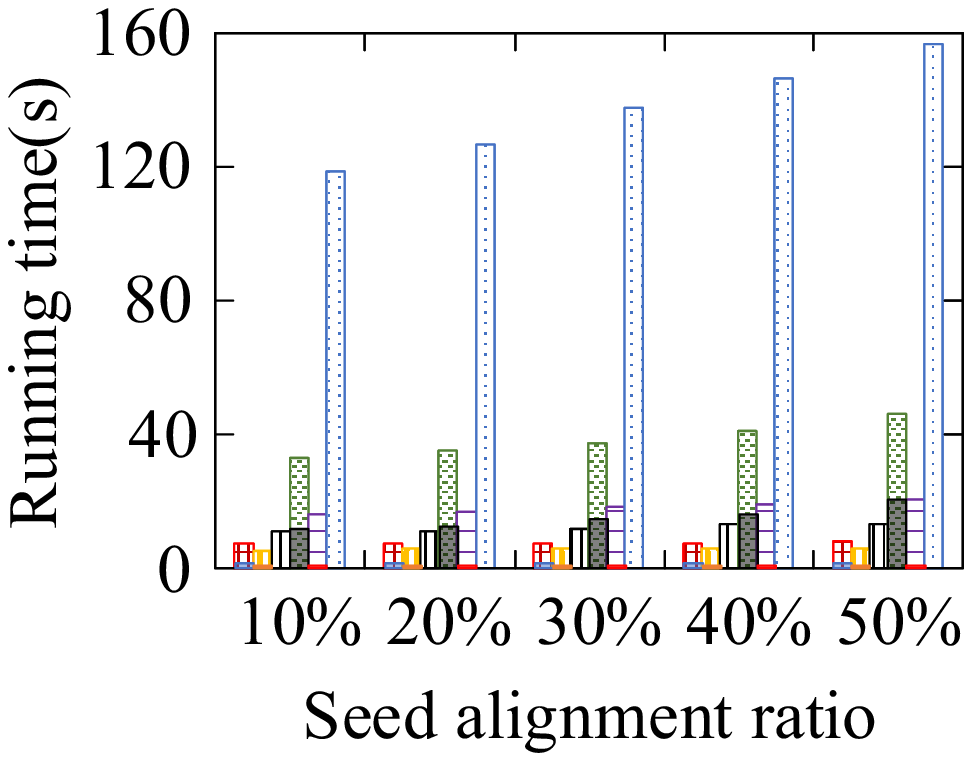}
}
\vspace*{-4mm}
\caption{METIS-CPS performance vs. seed alignment
}
\label{fig:before_after_exp}
\vspace*{-6mm}
\end{figure}

We explore the effect of the amount of seed alignment on METIS-CPS and VPS.
We report the H@1 and the running time produced by solely using the structural channel for EA when varying the seed alignment ratio from 10\% to 50\%.
Figure~\ref{fig:before_after_exp}(a) and Figure~\ref{fig:before_after_exp}(b) report the results of the alignment accuracy.
First, we observe that the H@1 of both METIS-CPS and VPS improves almost linearly as the number of seed alignment increases.
It is natural that more seed alignment provides more informative training signals to learn a reliable EA model.
Second, it is observed that METIS-CPS performs consistently much better than VPS no matter how the number of seed alignment changes.
This is because, METIS-CPS ensures that the more the seed alignment, the less destruction the KG's structure. Therefore, more structure features can be reserved to improve the EA performance produced by the structural channel.
On the contrary, VPS mainly relies on random partitioning, which greatly destroys the structure of a KG and thus results in minor accuracy increase.
Figure~\ref{fig:before_after_exp}(c) and Figure~\ref{fig:before_after_exp}(d) show the results of the running time required.
As expected, the running time of VPS is shorter than that of METIS-CPS.
This is because, the time complexity of VPS is lower than that of METIS-CPS, as mentioned in Section~\ref{sec:structure_channel}.
Although VPS is faster, it greatly destroys the structure of a KG.
We want to highlight that using METIS-CPS as the partition strategy can help the structural channel obtain higher alignment accuracy, which is much more critical for the task of EA.

\begin{table}[t]\small
\caption{Unsupervised EA results on DBP1M
}\label{exp:unsup}
\vspace*{-1.5mm}
\setlength{\tabcolsep}{0.1mm}{
\begin{tabular}{l|ccccc|ccccc}
\toprule
\multicolumn{1}{c|}{\multirow{2}{*}{Methods}} & \multicolumn{5}{c|}{DBP1M$_{EN-FR}$}  & \multicolumn{5}{c}{DBP1M$_{EN-DE}$}  \\ \cline{2-11}
\multicolumn{1}{c|}{} & \multicolumn{1}{c}{H@1} & \multicolumn{1}{c}{H@5} & \multicolumn{1}{c}{MRR} & \multicolumn{1}{c}{Time} & \multicolumn{1}{c|}{Mem.} & \multicolumn{1}{c}{H@1} & \multicolumn{1}{c}{H@5} & \multicolumn{1}{c}{MRR} & \multicolumn{1}{c}{Time} &\multicolumn{1}{c}{Mem.} \\ \hline
\textsf{LargeEA-G}$_{\mathbf{EN}\rightarrow \mathbb{L}}$ & 51.8 & 58.3 & 0.55 & 3.39 & 8.00G & 55.3 & 60.8 & 0.58 & 2.59 &4.59G\\
\textsf{LargeEA-G}$_{\mathbb{L}\rightarrow \mathbf{EN}}$ & 50.6 & 56.5 & 0.53 & 3.40 & 8.00G  & 55.6 & 61.3 & 0.58 & 2.64 & 5.37G \\
\textsf{LargeEA-R}$_{\mathbf{EN}\rightarrow \mathbb{L}}$ & 52.8 & 58.7 & 0.56 & 3.61 & 21.17G & 56.1 & 61.3 & 0.59 & 2.87 & 21.17G\\
\textsf{LargeEA-R}$_{\mathbb{L}\rightarrow \mathbf{EN}}$ &  51.5 & 57.0 & 0.54 & 3.72 & 21.17G  & 56.2 & 61.8 & 0.59 & 2.93 & 16.01G\\
\bottomrule
\end{tabular}}
\vspace*{-4mm}
\end{table}

Besides, one may be curious about the results of alignment accuracy and running time before and after the mini-batch generation.
In terms of \emph{accuracy}, as expected, the alignment accuracy after partition (w.r.t. METIS-CPS and VPS) is inferior to that before partition (w/o p. for short). This is because, graph partition inevitably cuts KG's edges and destroys the structure of a KG, thereby reducing the alignment accuracy.
In terms of \emph{running time},
it is observed that the running time of the structural channel with METIS-CPS is much shorter than that without partition.
The reason is that the subgraphs within each mini-batch (after performing METIS-CPS) are much smaller than the entire KG without partition. This leads to shorter training time for learning a reliable structure-based EA model in the structural channel, compared to the training process of the structural channel without partition.
To sum up, we want to emphasize the effectiveness of METIS-CPS for KG partition from the following two perspectives:
(i) \emph{Acceptable accuracy decline.} The average drop of alignment accuracy is around 8\% by performing METIS-CPS, compared against that without partition.
(ii) \emph{Faster training process.} Compared with the training process without partition, performing METIS-CPS can save up to 4x training time.

\vspace*{-1mm}
\subsection{Case Study: Unsupervised EA Performance}
%
To demonstrate the superior performance of \textsf{LargeEA} even when seed alignment is \emph{not} available, we present a case of applying \textsf{LargeEA} to conduct \emph{unsupervised} EA on DBP1M by using the proposed data augmentation strategy to generate seed alignment automatically.
Specifically, the data augmentation automatically generates 528,040 and 476,527 seeds on DBP1M$_{EN-FR}$ and DBP1M$_{EN-DE}$ respectively, with the accuracy of $93.86\%$ and $93.85\%$, respectively.
In this way, the proposed data augmentation can automatically generate sufficient high-quality seed alignment for EA.
Table~\ref{exp:unsup} reports the corresponding EA results.
We can observe that \textsf{LargeEA} is able to achieve an accuracy that is comparable with that under supervised EA.
This reflects that the name-based data augmentation can produce reliable pseudo seeds, which are able to provide positive input as real seeds.
This further confirms the superiority of the proposed \textsf{LargeEA} for coping with real-world EA scenarios.

\section{Related Work}
\label{sec:related_work}

Early EA methods rely on hand-crafted features~\cite{YAGO3}, crowdsourcing~\cite{Wikidata14, Hike17}, and OWL semantics~\cite{LogMap11}.
They are unrealistic for real-world EA scenarios with symbolic or linguistic heterogeneity.
Current EA approaches find equivalent entities by measuring the similarity between the embeddings of entities.

Structures of KGs are the basis for the embedding-based EA methods.
Representative EA approaches that purely rely on KGs' structures can be clustered into two categories, namely \emph{Translational}-\emph{based EA}~\cite{MTransE17, AKE19, SEA19, OTEA19, AttrE19, IPTransE17, BootEA18, TransEdge19} and \emph{GNN-based EA}~\cite{GCN-Align18, NAEA19, KECG19, MRAEA2020, RDGCN19, AliNet20, HyperKA20, MuGNN19}.
The former incorporates the translational KG embedding models (e.g., TransE~\cite{TransE13}) to learn entity embeddings; the latter
learns the entity embeddings by aggregating the neighbors' information of entities.
Though GNN-based models have demonstrated their outstanding performance, 
they suffer from poor scalability as they highly rely on the structure of KG, as mentioned in Section~\ref{sec:intro}. Therefore, \textsf{LargeEA} is developed to scale up these EA methods to align entities between large KGs.

Besides, lots of approaches have revealed that \emph{side information} of KGs can facilitate the EA performance,  including
\emph{entity names} \cite{JAPE17,GMNN19,MultiKE19,HGCN19,HMAN19,MRAEA2020,Coordinated20,DGMC20,DegreeAware20,NMN20,BERT-INT20,SSP20,RREA20,AttrGNN20}, \emph{descriptions} \cite{KDCoE18, HMAN19, MultiKE19, BERT-INT20}, \emph{images} \cite{EVA20}, and \emph{attributes} \cite{JAPE17,GCN-Align18,AttrE19,MultiKE19,HMAN19,COTSAE20,BERT-INT20,AttrGNN20,EPEA20}.
They could be considered as complements to, but not competitors of, the structure-based EA models.
Since every entity has its own name and the use of name information does not require any pre-processing, users/researchers tend to use the name information to promote EA.
Other studies also reveal that entities' descriptions, attributes, and images contain more information than entities' name.
However, these studies are either labor-intensive or error-prone and thus restrict the scope of their real-world applications.
To this end, we incorporate name information in \textsf{LargeEA}.



\section{Conclusions}
In this paper, we present \textsf{LargeEA} to align entities between large-scale knowledge graphs.
\textsf{LargeEA} introduces both \emph{structure channel} and \emph{name channel} to collaboratively align entities from large-scale KGs.
In the structure channel, we propose METIS-CPS to generate multiple mini-batches and then learn the structural features of entities within each mini-batch independently.
In the name channel, we explore the name features of entities from both string-based aspect and semantic aspect without any complex training process via our proposed NFF.
Additionally, we exploit a name-based data augmentation to enrich the seed alignment for EA.
The EA results of \textsf{LargeEA} are derived from the fused features of names and structures.
To simulate real-world EA scenarios, we also develop a large-scale EA benchmark named DBP1M for evaluating EA performance.
Considerable experimental results on EA benchmarks with different data magnitudes demonstrate the superiority of \textsf{LargeEA}.
In the future, we would like to explore scalable EA approaches that only rely on the KG's structure, to support EA between KGs whose entities do not share the same naming convention.

\label{sec:conclusions}


\bibliographystyle{ACM-Reference-Format}
\bibliography{refer}

\begin{appendices}

\section{the internal functionality of structure channel}

To illustrate how to compute the structure-based entity similarities via the structure channel clearer, we present the pseudo-code in Algorithm~\ref{algorithm:sc} to outline the internal functionality of the structure channel.

\begin{algorithm}[h]
\LinesNumbered
\DontPrintSemicolon
\caption{Computing the structure-based entity similarities via structural channel}
\label{algorithm:sc}
    \KwIn{a source KG $G_s$ and a target KG $G_t$}
    \KwOut{the structural feature matrix $\bm{M}_s$}
    
    getting pseudo seeds $\psi'_p$ by name-based data augmentation\;
    
    $\psi' \leftarrow \psi' + \psi'_p$\;
    
    $\{B_1, B_2, \cdots, B_K\} \leftarrow$ mini-batch-generation($G_s$, $G_t$, $\psi'$)\;
    
    \ForEach{mini-batch $B_i$}{
        $\bm{H}_{s}^{i}, \bm{H}_{t}^{i'} \leftarrow $  mini-batch-training($G_s^i \in B_i$, $G_t^i \in B_i$)\;
    }
    $\bm{M}_s \leftarrow$ calculating Manhattan distance between each entity pair $(\bm{h}_{e_s^i}, \bm{h}_{e_{t}^{i'}})$\;
\Return{$\bm{M}_s$}
\end{algorithm}

The structural channel takes as input a source KG $G_s$ and a target KG $G_t$. First, it produces a set of pseudo seeds $\psi'_p$ via the name-based data augmentation to enrich the seed alignment (Lines 1-2). Second, it generates $K$ mini-batches according to the proposed mini-batch generation strategies, i.e., METIS-CPS or VPS, as described in Section~\ref{sec:mini-batch-generation} (Line 3).
Then, for each mini-batch $B_i$ (w.r.t. $G^i_s$ and $G^i_t$), it applies a GNN-based EA model to train the embeddings of entities, denoted as $\bm{H}_{s}^{i}$ and  $\bm{H}_{t}^{i'}$. The former represents the set of entity embeddings from $G_s^{i}$; while the latter is the set of entity embeddings from $G_t^{i}$ (Lines 4-5).
Finally, it calculates the Manhattan distance between each entity pair ($\bm{h}_{e_s^i}$, $\bm{h}_{e_{t}^{i'}}$) to form the structure-based entity similarity matrix $M_s$, where $\bm{h}_{e_s^i} \in \bm{H}_{s}^{i}$ and $\bm{h}_{e_{t}^{i'}} \in \bm{H}_{t}^{i'}$ (Lines 6-7).

\section{Mini-batch Number Effect}

\begin{figure}[!]
\centering
\includegraphics[width=0.5\textwidth]{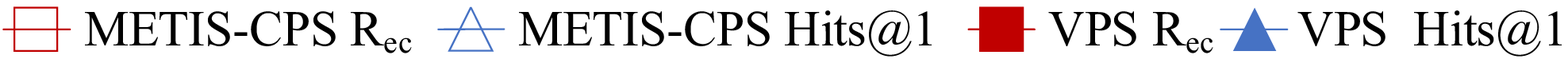}
\vspace*{-5mm}\\
\subfigure[EN-FR]{
    \includegraphics[width=1.68in]{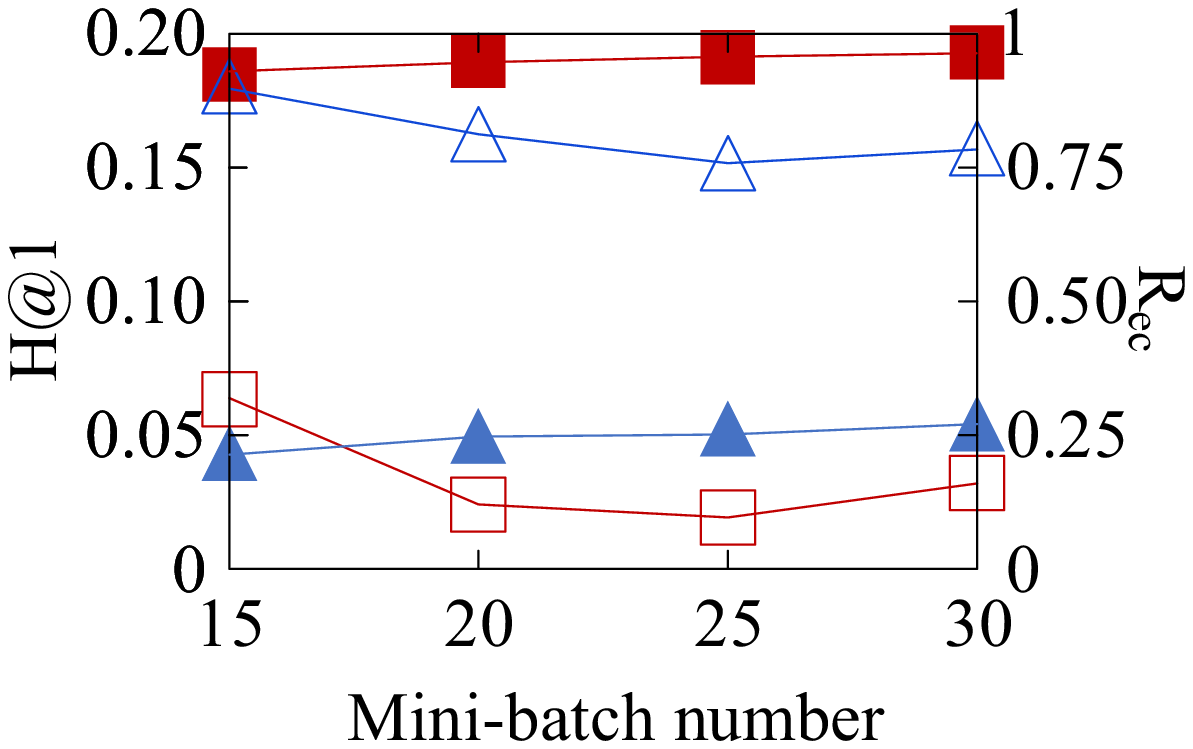}
}\hspace{-5mm}
\subfigure[EN-DE]{
 \includegraphics[width=1.68in]{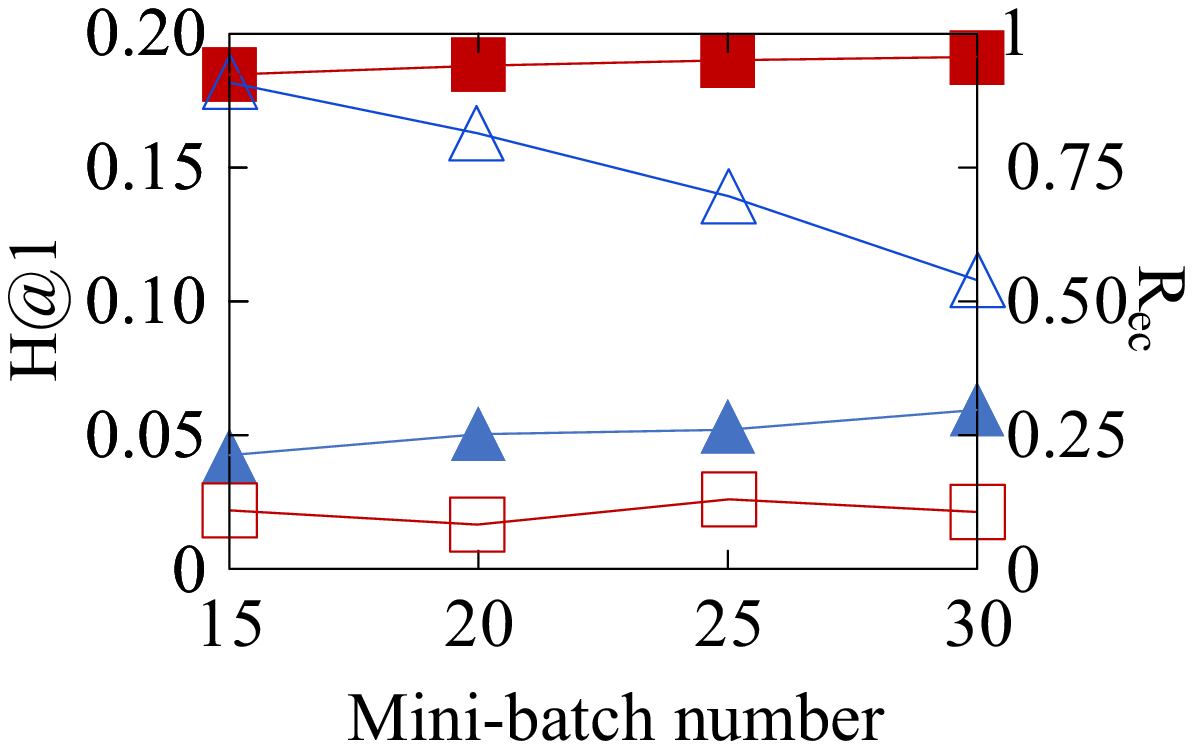}
}
\caption{Mini-batch Generation vs. Mini-batch Number}
\label{fig:mini_batch_exp}
\end{figure}

We first compare the performance of METIS-CPS and VPS on DBP1M by varying the mini-batch number $K$ from 15 to 30.
Note that $K=15$ is the minimal mini-batch number that allows \textsf{LargeEA} to work on datasets in DBP1M in the experimental conditions.
Since varying $K$ only affects the structure channel, we only report the accuracy of EA results (w.r.t. $H@1$) produced by the structure channel (i.e., \textsf{LargeEA} w/o name channel).
We try to quantify the loss of KGs' structures and introduce 
a new metric \emph{edge-cut rate} ($R_{ec}$ for short) to assist the analysis of mini-batch generation performance.
$R_{ec}$ is defined as the fraction of the number of edges/triples cut off to the total number of edges in the original KGs.
The results are shown in Figure~\ref{fig:mini_batch_exp}.

The first observation is that the accuracy of METIS-CPS drops as the mini-batch number $K$ increases.
This is because the larger the mini-batch number, the more edge-cuts in each KGs, and the more difficult to learn reliable structural features from an EA model.
The second observation is that no matter how $K$ changes, METIS-CPS always outperforms VPS on H@1.
The reason is that VPS divides each KG by randomly deleting edges and hence significantly destroys the structure of a KG; while METIS-CPS is able to preserve the structural features much better.
The third observation is that, the accuracy of METIS-CPS drops more slowly on DBP1M$_{EN-FR}$ datasets than its accuracy on DBP1M$_{EN-DE}$ datasets.
This is because German KGs are relatively sparser than French KGs in every benchmark, and they become much sparser when $K$ ascends.
Therefore, it is much harder for an EA model to learn effective structural features from a sparser KG than from a relatively denser KG.
Overall, METIS-CPS consistently outperforms VPS, which 
demonstrates the effectiveness of METIS-CPS.

\section{Overlapping Mini-batches Analysis}

\begin{table*}[t]
\caption{Percentage of Equivalent Entities that are Placed into the same Mini-batches (The Best Scores are Marked in \textbf{Bold})
}\label{exp:partition_quality}
\setlength{\tabcolsep}{3mm}{
\begin{tabular}{c|c|cc|cc|cc}
\toprule
\multirow{2}{*}{Dataset}    & \multirow{2}{*}{Mini-batch generation method} & \multicolumn{2}{c|}{Total}        & \multicolumn{2}{c|}{Training set}  & \multicolumn{2}{c}{Test set}   \\ \cline{3-8} 
    & & EN$\rightarrow \mathbb{L}$ & $\mathbb{L}\rightarrow$EN & EN$\rightarrow \mathbb{L}$ & $\mathbb{L} \rightarrow$EN & EN$\rightarrow \mathbb{L}$ & $\mathbb{L} \rightarrow$EN \\ \hline
\multirow{2}{*}{IDS15K$_{EN-FR}$}  & METIS-CPS   & \textbf{64.9}         & \textbf{61.3}         & 76.4      & 68.1      & \textbf{62.0}         & \textbf{59.9}         \\
    & VPS         & 36.1      & 36.1      & \textbf{100.0}        & \textbf{100.0}        & 20.7      & 20.7      \\ \hline
\multirow{2}{*}{IDS15K$_{EN-DE}$}  & METIS-CPS   & \textbf{86.2}         & \textbf{86.2}         & 92.2      & 92.3      & \textbf{84.7}         & \textbf{84.7}         \\
    & VPS         & 36.4      & 36.4      & \textbf{100.0}        & \textbf{100.0}        & 20.7      & 20.7      \\ \hline
\multirow{2}{*}{IDS100K$_{EN-FR}$} & METIS-CPS   & \textbf{60.8}         & \textbf{59.1}         & 74.7      & 66.6      & \textbf{57.4}         & \textbf{57.3}         \\
    & VPS         & 27.9      & 27.9      & \textbf{100.0}        & \textbf{100.0}        & 10.6      & 10.5      \\ \hline
\multirow{2}{*}{IDS100K$_{EN-DE}$} & METIS-CPS   & \textbf{61.4}         & \textbf{60.4}         & 75.1      & 71.1      & \textbf{58.3}         & \textbf{57.8}         \\
    & VPS         & 28.0      & 28.0      & \textbf{100.0}        & \textbf{100.0}        & 10.5      & 10.5      \\ \hline
\multirow{2}{*}{DBP1M$_{EN-FR}$}   & METIS-CPS   & \textbf{42.0}         & \textbf{52.5}         & 76.3      & 71.5      & \textbf{30.3}         & \textbf{45.1}         \\
    & VPS         & 24.0      & 24.2      & \textbf{100.0}        & \textbf{100.0}        & 5.2       & 5.2       \\ \hline
\multirow{2}{*}{DBP1M$_{EN-DE}$}   & METIS-CPS   & \textbf{46.7}         & \textbf{54.9}         & 83.5      & 80.5      & \textbf{32.7}         & \textbf{45.2}         \\
    & VPS         & 26.8      & 25.4      & \textbf{100.0}        & \textbf{100.0}        & 6.5       & 5.9       \\ \bottomrule
\end{tabular}}
\end{table*}

Since the mini-batches generated by METIS-CPS are disjoint with zero overlap,
one may wonder whether overlapping mini-batches can affect the performance of \textsf{LargeEA} (with METIS-CPS for mini-batch generation).
We introduce parameter $D_{ov}$ to determine how the overlapping mini-batches are generated. Specifically, given in total $K$ mini-batches, each mini-batch is combined with top-$k$ most similar mini-batches to form an overlapping mini-batch. The $K$ mini-batches are then replaced by the $K$ overlapping mini-batches generated. Note that, when $D_{ov} = 1$, each mini-batch is combined with itself, the top-1 most similar mini-batch to itself, so the generated mini-batches are still disjoint; when $D_{ov} = 3$, each mini-batch is combined with another two mini-batches that are most similar to it. 

\begin{figure}[t]
\centering
\subfigure[EN-FR]{
    \includegraphics[width=1.7in]{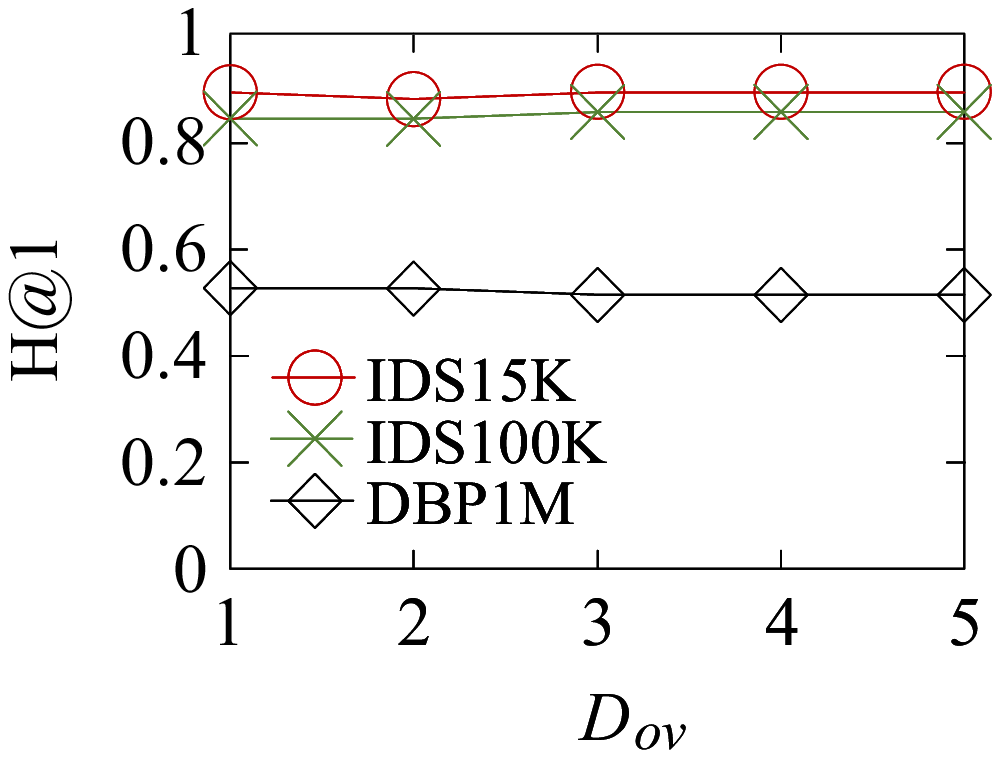}
}\hspace{-6mm}
\subfigure[EN-DE]{
 \includegraphics[width=1.7in]{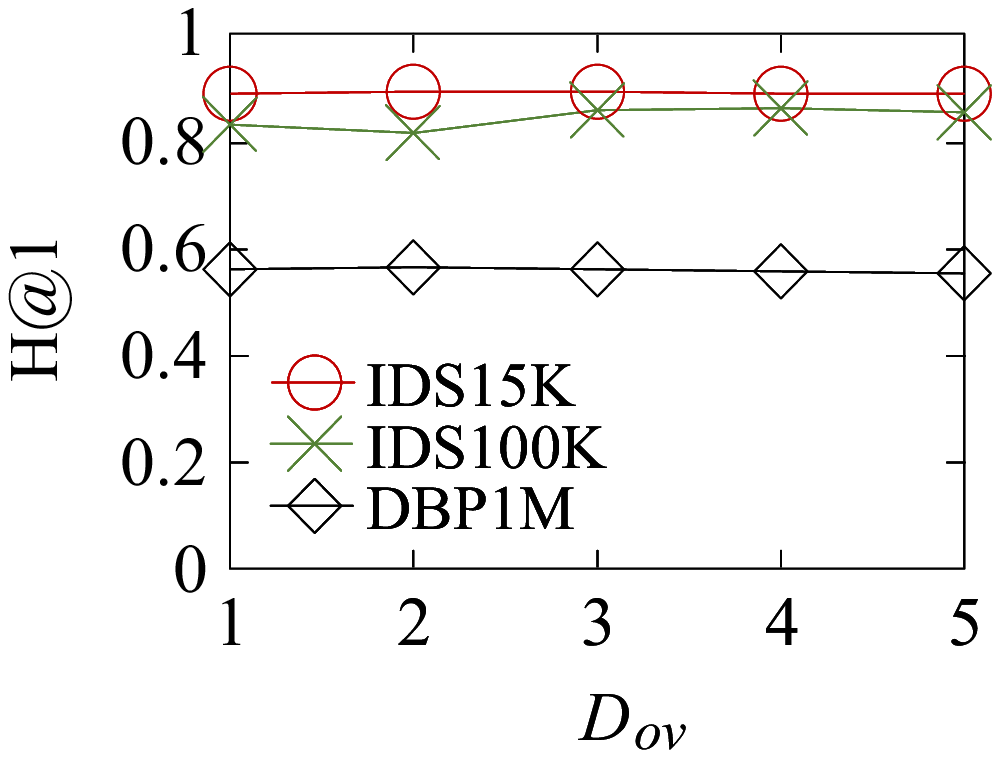}
}
\vspace*{-2mm}
\caption{Mini-batch Generation vs. Overlapping}
\label{fig:overlap_batch_exp}
\end{figure}

The results are shown in Figure~\ref{fig:overlap_batch_exp}.
It is observed that the EA accuracy of structure channel (i.e, H@1) remains almost unchanged when $D_{ov}$ changes its values. 
This is because, although increasing the degree of overlap enables more equivalent entities to be found within a mini-batch, it also introduces more invalid candidates that cannot be aligned in the mini-batch. That's why we propose the design of non-overlapping mini-batches, which introduces two benefits.  
First, it eliminates expensive communication with other mini-batches.
Second, it reduces the number of entities and relations involved in every mini-batch, resulting in a smaller memory footprint of GPU.

\section{Partition Quality Analysis}

As entities from different mini-batches are not able to be aligned, it's critical to ensure that our mini-batch generation algorithm is able to place equivalent entities into the same mini-batch. Accordingly, we evaluate the number of equivalent entities that are placed into the same mini-batch by adopting METIS-CPS.
It is challenging to theoretically ensure that our partitioning algorithms are able to place all the equivalent entities into the same mini-batch. However, in our experimental study, the proposed METIS-CPS is able to place the considerable amount of the equivalent entities into the same mini-batch. Table~\ref{exp:partition_quality} reports the corresponding results.

For the training set, as can be observed, the performance of METIS-CPS is inferior to that of VPS for preserving equivalent entities into the same mini-batch.
This is due to the inherent characteristics of these methods.
For VPS, it is designed to put the pre-defined equivalent entities (i.e., seed alignment) into the same mini-batch.
Thus, as expected, 100\% equivalent entities in the training set are correctly placed into the same mini-batch.
However, METIS-CPS is designed to not only put seed alignment into the same mini-batch as much as possible but also preserve the structural properties of KGs as much as possible. 
Recall that it is challenging to achieve the above two goals simultaneously, leading to the inferior performance of METIS-CPS in the training set.

For the test set, we observe that METIS-CPS achieves superior performance than VPS on all datasets.
This verifies the effectiveness of METIS-CPS in preserving unknown equivalent entities into the same mini-batch.
Specifically, recall that many studies~\cite{RREA20, AliNet20, GCN-Align18} have illustrated that the neighbors of two known equivalent entities are likely to be equivalent.
In view of this, METIS-CPS can put unknown equivalent entities into the same mini-batch to a certain extent according to both the guidance of the seed alignment and the goal of structure preservation.
On the contrary, VPS purely relies on the random partition and hence there is no guarantee that unknown equivalent entities (in the test set) can be put into the same mini-batch.

We would like to highlight that putting more unknown equivalent entities in the test set is much more important than that in the training set. This is because the purpose of entity alignment is to align unknown equivalent entities correctly. If two unknown equivalent entities are placed into different mini-batches, they can not be aligned to any further extent. In addition, it is observed that the corresponding results of DBP1M are relatively inferior to the results of IDS. This is because IDS contains more structure features than DBP1M (as discussed in Section~\ref{sec:ablation}), which contribute to better KG partition results.

\section{The GPU Memory Usage of \textsf{LargeEA}}

\begin{table}[t]
\caption{The GPU Memory Usage of LargeEA}
\label{exp:memory1}
\setlength{\tabcolsep}{0.5mm}{
\begin{tabular}{c|c|c|cc}
\toprule
\multirow{2}{*}{Dataset} & \multirow{2}{*}{Settings} & \multirow{2}{*}{Name Channel} & \multicolumn{2}{c}{Structure Channel} \\ \cline{4-5} 
 &   &       & \textsf{LargeEA-R}          & \textsf{LargeEA-G}         \\ \hline
\multirow{4}{*}{IDS15K}  & EN$\rightarrow$FR         & 1.54G & 1.02G/4.07G & 0.25G/1.00G \\
 & FR$\rightarrow$EN         & 1.54G & 1.01G/4.07G  & 0.25G/0.50G \\
 & EN$\rightarrow$DE         & 1.54G & 1.01G/4.07G  & 0.25G/1.00G \\
 & DE$\rightarrow$EN         & 1.54G & 1.01G/4.07G  & 0.25G/0.50G \\ \hline
\multirow{4}{*}{IDS100K} & EN$\rightarrow$FR         & 1.74G & 4.04G/--  & 1.06G/4.00G \\
 & FR$\rightarrow$EN         & 1.74G & 4.04G/--  & 1.00G/4.00G\\
 & EN$\rightarrow$DE         & 1.74G & 4.04G/--  & 1.00G/4.00G \\
 & DE$\rightarrow$EN         & 1.74G & 4.04G/--  & 0.50G/4.00G \\ \hline
\multirow{4}{*}{DBP1M}   & EN$\rightarrow$FR         & 6.61G & 21.15G/-- & 4.00G/-- \\
 & FR$\rightarrow$EN         & 7.37G & 21.17G/-- & 8.00G/-- \\
 & EN$\rightarrow$DE         & 4.59G & 16.01G/-- & 4.00G/-- \\
 & DE$\rightarrow$EN         & 5.36G & 16.01G/-- & 4.00G/-- \\
\bottomrule
\end{tabular}}
\vspace{-4mm}
\end{table}

We also report the GPU memory usage of the proposed \textsf{LargeEA} in Table~\ref{exp:memory1} for further investigation, where the symbol slash (/) separates the left values from the right values.
The left ones represent the results when METIS-CPS is used in the structure channel, while the right ones correspond to results without partition.
It is observed that performing structure channel usually requires more GPU memory on large dataset (i.e., DBP1M), compared to the name channel.
This is because the structure channel contains a complex EA training process, which consumes a large amount of GPU memory for storing the parameters used in EA training.
Differently, the name channel does not require any training process, resulting in a relatively smaller usage of GPU memory.
We also observe that applying the mini-batch generation strategy  significantly reduces the usage of GPU memory in the structure channel, compared to that without performing any partition.
Recall that obtaining effective mini-batch generation results is challenging, as mentioned in Section~\ref{sec:intro}.
The results confirm the necessity of utilizing METIS-CPS for saving the GPU memory.

\end{appendices}

\balance

\end{document}